\documentclass[conference,compsoc]{IEEEtran}
\usepackage{amsmath,amssymb}        
\usepackage{url}                    
\usepackage{etoolbox}               

\usepackage{tabularx}               
\usepackage{booktabs}               
\usepackage{multirow}               

\usepackage{minted}                  
\usepackage{listings}
\usepackage{caption}
\setminted{
    fontsize=\footnotesize,
    linenos,
    breaklines,
    frame=lines,
    numbers=none,
    framesep=1.5mm,
    rulecolor=\color{black},
    texcomments
}

\lstdefinelanguage{diff}{
  language=Python,
  morecomment=[f][\color{diffaddcolor}]{+},
  morecomment=[f][\color{diffdelcolor}]{-},
  morecomment=[f][\color{gray}]{@},
  escapeinside={|}{|}
}

\lstdefinestyle{pythonstyle}{
    language=diff,
    basicstyle=\ttfamily\footnotesize,
    keywordstyle=\color{codegreen}\bfseries,
    commentstyle=\color{mintedcomment}\itshape,
    stringstyle=\color{mintedstring},
    showstringspaces=false,
    frame=lines,
    numbers=none,
    breaklines=true,
    tabsize=4
}

\usepackage{algorithm}
\usepackage{algpseudocode}

\algnewcommand{\LeftComment}[1]{\small\texttt{\Statex \(//\) #1}}

\usepackage{tcolorbox}               
\usepackage{soul}                    
\setstcolor{violet}

\usepackage{enumitem}                
\usepackage{url}                     

\usepackage{xcolor}

\definecolor{codegreen}{rgb}{0,0.6,0}
\definecolor{codeblue}{rgb}{0,0,0.6}
\definecolor{conclusiongray}{gray}{0.92}
\definecolor{githubgreen}{RGB}{230,255,237}
\definecolor{githubredlight}{RGB}{255,235,233}
\definecolor{diffaddcolor}{rgb}{0.1, 0.7, 0.1}
\definecolor{diffdelcolor}{rgb}{0.7, 0.1, 0.1}
\definecolor{slightgray}{HTML}{A9A9A9}
\definecolor{slightorange}{HTML}{FF8C69}
\definecolor{slightblue}{HTML}{87CEEB}
\definecolor{commentcolor}{rgb}{0.25, 0.5, 0.25}
\definecolor{stringcolor}{rgb}{0.1, 0.1, 0.9}
\definecolor{base}{HTML}{8d918b}
\definecolor{adv}{HTML}{b3cfac}
\definecolor{mustard}{HTML}{eabc31}
\definecolor{mintedcomment}{rgb}{0.0, 0.0, 0.75}
\definecolor{mintedkeyword}{rgb}{0.3, 0.4, 0.3}
\definecolor{mintedstring}{rgb}{0.55, 0.2, 0.1}

\usepackage{tikz}                    
\usetikzlibrary{patterns}
\usetikzlibrary{patterns,patterns.meta}

\definecolor{BaseBlue}{HTML}{0072B2}
\definecolor{GraspOrange}{HTML}{E69F00}
\definecolor{lightGreen}{HTML}{6BAF92}

\DeclareRobustCommand{\legendrect}[4]{%
  \tikz[baseline=0.1ex]{%
    \shorthandoff{:;}%
    \fill[#3] (0,0) rectangle (#2,#1);
    \draw[pattern={#4}, pattern color=black] (0,0) rectangle (#2,#1);
  }%
}

\providecommand{\shorthandoff}[1]{}

\newcommand{\yes}{\tikz\draw[fill=black] (0,0) circle (0.6ex);}
\newcommand{\no}{\tikz\draw[draw=black,fill=white] (0,0) circle (0.6ex);}
\newcommand{\half}{\tikz{
    \draw[draw=black,fill=white] (0,0) circle (0.6ex);
    \clip (-0.6ex,-0.6ex) rectangle (0,0.6ex);
    \fill[black] (0,0) circle (0.6ex);}
}

\newcommand{\ProjectName}[1]{{\small\textsc{GRASP}}}
\newcommand{\SecProjectName}[1]{{\large\textsc{\textbf{GRASP}}}}
\newcommand{\TabProjectName}[1]{{\textsc{GRASP}}}

\usepackage{hyperref}
\usepackage[T1]{fontenc}
\ifCLASSOPTIONcompsoc
  \usepackage[nocompress]{cite}
\else
  \usepackage{cite}
\fi
\ifCLASSINFOpdf
\else
\fi
\usepackage{array}

\begin{document}

\date{} 

\title{Fortifying LLM-Based Code Generation with Graph-Based Reasoning on\\ Secure Coding Practices}

\author{
\IEEEauthorblockN{
Rupam Patir,
Keyan Guo,
Haipeng Cai,
Hongxin Hu
}
\IEEEauthorblockA{
University at Buffalo\\
\{rupampat, keyanguo, haipengc, hongxinh\}@buffalo.edu
}
}

\date{} 

\maketitle

\begin{abstract}
The code generation capabilities of Large Language Models (LLMs) have transformed the field of software development. However, this advancement also presents significant security challenges, as LLM-generated code often contains vulnerabilities. One direction of research strengthens LLMs by injecting or refining security knowledge through curated datasets, model tuning, or static analyzers. While effective in certain settings, these methods can be resource-intensive, less adaptable to zero-day vulnerabilities, and often inapplicable to proprietary models. To address these challenges, we introduce \ProjectName{}, which explores a new direction that focuses on structured reasoning over Secure Coding Practices~(SCPs) rather than additional training or external feedback. \ProjectName{} comprises two key ideas: (1) an SCP graph that organizes SCPs into a Directed Acyclic Graph (DAG) capturing dependencies and relationships, and (2) a graph-based reasoning process that systematically guides LLMs through relevant SCPs for code generation. This design enables interpretable, model-agnostic, and scalable security improvements, particularly for previously unseen vulnerabilities. Our evaluation shows that \ProjectName{} consistently achieves Security Rates (SR) exceeding 80\% across multiple LLMs, and delivers up to 88\% improvements over baselines on zero-day vulnerabilities.
\end{abstract}
\section{Introduction}

The rapid advancements in Large Language Models~(LLMs) have transformed software development by enhancing coding capabilities~\cite{openai, claude, deepmind2024gemini, fried2023incodergenerativemodelcode, Li_2022, li2023starcodersourceyou}. 
Research highlights the productivity benefits of these models in software development, with GitHub Copilot users completing tasks 55\% faster, experiencing improved job satisfaction, and reduced cognitive load~\cite{github_copilot, github_copilot_impact}. Google's findings further highlight the advantages of ML-enhanced code completion, demonstrating a 6\% reduction in time between builds and tests with single-line ML completion, along with a 25-34\% user acceptance rate for code suggestions~\cite{google_ml_code_completion_2024}.

Despite these advancements, critical security concerns persist in adopting LLMs for code generation. Studies have shown that LLM-generated code frequently contains vulnerabilities. 
For instance, GitHub Copilot was found to produce vulnerable code in approximately 40\% of cases~\cite{pearce2021asleepkeyboardassessingsecurity, asare2024githubscopilotbadhumans}. 
More recently, Mou et al.~\cite{mou2025reallytrustcodecopilots} reported that popular LLMs such as GPT-4o generate secure code only 70–75\% of the time. 
These rates are untenable in practice, as insecure code can readily lead to data breaches, privilege escalation, or widespread system compromise.

To mitigate these concerns, researchers have proposed several techniques to strengthen LLMs for secure code generation. SVEN~\cite{He_2023} applied prefix tuning to improve generation security. SafeCoder~\cite{he2024instructiontuningsecurecode} extended this approach using instruction tuning to provide more flexible control over secure code generation.
PromSec~\cite{nazzal2024promsec} introduced prompt-level optimization through graph-based adversarial training, leveraging curated vulnerability datasets and external static analyzers such as Bandit~\cite{bandit} and SpotBugs~\cite{spotbugs} to guide the refinement process. Together, these methods represent one direction of research that strengthens LLMs by equipping them with additional security knowledge or refining it through external signals. While effective, this direction faces persistent challenges, including reliance on curated datasets that miss zero-day vulnerabilities, the need for model access unavailable in proprietary systems, added overhead from external analyzers and limited interpretability.

In this paper, we pursue a novel direction that focuses on ensuring the reliable and systematic operationalization of security knowledge.
Recent work shows that LLMs already exhibit awareness of security concepts relevant to software development~\cite{ullah2024llmsreliablyidentifyreason, fu2024constraineddecodingsecurecode}. In this case, the challenge is not acquiring knowledge but ensuring its consistent application during code generation. This difficulty mirrors a longstanding issue in human software development and reflects the well-documented gap between \emph{security theory} and \emph{coding practice}\cite{scpsinjava}. Developers may understand secure design concepts abstractly but fail to apply them under real-world conditions. To address these gaps, the software engineering community has long relied on Secure Coding Practices (SCPs)~\cite{owasp_secure_coding_guide, cert_secure_coding_standards, microsoft_secure_coding_guidelines}, which provide the structured discipline needed to translate abstract knowledge into consistent implementation. Empirical studies~\cite{scpsinjava} further confirm that adherence to SCPs enabled developers to produce more secure code in practice, motivating us to consider whether similar structured practices can help LLMs.

We therefore extend this precedent by adapting it to LLMs. Just as disciplined methods helped developers reliably apply what they already knew, structured mechanisms can guide LLMs in transforming their latent security knowledge into secure outputs. \ProjectName{} builds on this by operationalizing SCPs for LLM-based code generation. It does so through two key ideas: (i) an SCP Graph that organizes practices into a Directed Acyclic Graph (DAG) encoding their dependencies and relationships, and (ii) a graph-based reasoning process that systematically applies these practices in context. In this way, \ProjectName{} represents a novel direction relative to prior work, shifting the focus from augmenting security knowledge through retraining or curated datasets to ensuring its reliable and systematic operationalization. This design yields natural benefits such as generalization to zero-day vulnerabilities, model agnosticism, freedom from external analyzers, and improved interpretability.

The key contributions of this paper are as follows:
\begin{itemize}[left=0em]
    \item \textbf{We investigate the role of Secure Coding Practices (SCPs) in conditioning LLMs for secure code generation.} SCPs, widely used by developers to prevent vulnerabilities, can also be leveraged to guide LLMs toward more secure outputs. We advance this premise by modeling SCPs as a structured graph and analyzing their effectiveness for security-aware generation.
    
    \item \textbf{We propose \ProjectName{}\footnote[1]{We will make our data and code publicly available in advance of the paper's publication.}, a reasoning-driven framework that leverages an SCP Graph to fortify the code generation process.}
    The SCP Graph is a Directed Acyclic Graph that encodes dependency relationships among SCPs. \ProjectName{} traverses this graph dynamically based on task relevance, incrementally applying security transformations while maintaining logical consistency and functional correctness.
    
    \item \textbf{We construct a benchmark dataset for secure code generation, annotated with unit tests to enable joint evaluation of security and functionality.}
    The benchmark consists of a diverse collection of CWE-based scenarios, each paired with unit tests to ensure functional correctness.  To the best of our knowledge, this is the first benchmark for secure code generation that simultaneously accomplishes \textbf{both} (a) the use of natural language prompts instead of code completion prompts, and (b) the joint evaluation of security and functional correctness.
    
    \item \textbf{We conduct comprehensive experiments across multiple LLMs, CWEs, and zero-day CVEs.} \ProjectName{} improves the overall Security Rate (SR) to over \textbf{80\%} for Claude, GPT-4o, Gemini and Llama3, while maintaining functional correctness. It also generalizes to unseen vulnerabilities, achieving SR gains of up to \textbf{88\%} over baseline methods on real-world CVEs.
\end{itemize}

\section{Background and Related Work}

In this section, we provide the necessary background knowledge and discussion of closely related work.

\subsection{LLMs in Code Generation}

Large Language Models (LLMs) have significantly impacted the field of code generation, offering new possibilities for automating and assisting in software development tasks. Brown et al.~\cite{NEURIPS2020_1457c0d6} demonstrated the potential of large-scale language models to perform various tasks, including code generation, with minimal task-specific fine-tuning. Chen et al.~\cite{chen2021evaluatinglargelanguagemodels} further explored the capabilities of LLMs trained specifically in code, highlighting their potential and limitations. The development of CodeBERT by Feng et al.~\cite{feng2020codebertpretrainedmodelprogramming} showcased the effectiveness of pre-training models on both programming languages and natural language descriptions.

LLMs have emerged as prominent tools for software developers in code-generation tasks. 
Proprietary models such as OpenAI's GPT series~\cite{openai}, Anthropic's Claude~\cite{claude}, and Google's Gemini~\cite{deepmind2024gemini} offer advanced code generation features with strong instruction-following abilities. They are designed to understand and generate code across various programming languages. 
GitHub Copilot~\cite{github_copilot}, built on OpenAI's Codex~\cite{openai2024codex}, is specifically tailored for code completion and generation tasks within development environments. 
In contrast, open-source models like Meta's LLama3~\cite{meta2024llama} and Salesforce's CodeGen~\cite{nijkamp2023codegenopenlargelanguage} provide accessible alternatives. 
These models differ in their accessibility, initial training focus, and level of instruction tuning, with proprietary models often offering more advanced performance for code-related tasks~\cite{paperswithcode_humaneval_sota}.

\subsection{Vulnerable Code}

Vulnerable code encompasses software flaws that can be exploited to compromise the security of systems, potentially leading to significant issues such as unauthorized access and data breaches. Addressing these flaws is crucial for ensuring the security and integrity of software applications~\cite{6405650}.

The Common Weakness Enumeration (CWE) framework~\cite{mitre2024cwe}, maintained by the MITRE Corporation, categorizes these vulnerabilities into a structured list of weaknesses that can lead to security risks. As part of their efforts, MITRE maintains a list of the Top 25 CWEs, which highlights the most critical and prevalent weaknesses in software~\cite{mitre2023top25}. For example, SQL Injection (CWE-89) is a prevalent vulnerability that allows attackers to manipulate SQL queries through malicious input, potentially exposing or altering database content. Cross-site scripting (XSS) (CWE-79) is a vulnerability that allows attackers to inject malicious scripts into web applications, potentially leading to unauthorized access or data theft. Cross-Site Request Forgery (CSRF) (CWE-352) tricks users into performing unintended actions on a Web application. Insecure Deserialization (CWE-502) involves the unsafe handling of serialized data, which can be exploited to execute arbitrary code or alter application data.

The impact of these vulnerabilities can be substantial. Data breaches resulting from such flaws compromise user privacy, incur significant financial losses due to remediation efforts and downtime, and cause reputation damage that erodes customer trust~\cite{ibm2024databreach, ibm2024reputationalrisk}. Effective detection and mitigation of these vulnerabilities involves techniques such as static code analysis~\cite{codeql2024, sonarsource2024sonarqube, semgrep2024} and dynamic code analysis~\cite{zap2024, veracode2024dast, invicti2024}, complemented by manual code reviews.

\subsection{Secure Coding Practices}

Secure coding practices (SCPs) are essential for coaching software developers in the development of resilient and secure software. They encompass a variety of techniques designed to prevent vulnerabilities and ensure that the code behaves securely under various conditions~\cite{meng2017securecodingpracticesjava, 8511955}. For instance, to prevent SQL Injection, which occurs when malicious input manipulates SQL queries, it is crucial to use parameterized queries and prepared statements. This ensures that user input is treated as data rather than as executable code. Similarly, to mitigate cross-site scripting (XSS), it is vital to ensure that user input is properly sanitized before being rendered.

Resources such as OWASP Secure Coding Practices~\cite{owasp_secure_coding_guide}, CERT Secure Coding Standards~\cite{cert_secure_coding_standards}, and Microsoft Secure Coding Guidelines~\cite{microsoft_secure_coding_guidelines} provide comprehensive guidance, including centralized input validation, secure data encoding/decoding, and robust authentication and session management. Following these practices helps developers significantly reduce the risk of vulnerabilities.

\subsection{LLM-Based Code Generation Fortification}

Improving the security of LLM-based code generation has gained increasing attention in recent research. He et al.~\cite{He_2023} introduced SVEN, a control method that applies prefix-tuning to steer models toward generating secure or insecure code. Rather than updating the model’s original weights, this approach trains only lightweight prefix parameters, thereby reducing the number of trainable components while retaining flexibility for code generation tasks. Building on this direction, He et al.~\cite{he2024instructiontuningsecurecode} proposed SafeCoder, which performs security-focused instruction tuning using a large-scale dataset of verified vulnerability fixes gathered from GitHub commits. 
SafeCoder jointly optimizes for both security and functional utility, yielding a notable 30\% improvement in code security across diverse tasks. However, such approaches require considerable computational resources and may overfit to specific vulnerability patterns. Beyond model tuning, prompt engineering has been leveraged to craft prompts that encourage secure coding~\cite{res2024enhancingsecurityaibasedcode}. 

Recent work by PromSec~\cite{nazzal2024promsec} advances this paradigm by introducing a generative adversarial graph neural network (gGAN)-based framework to iteratively optimize prompts for secure code generation. PromSec employs a dual-objective contrastive learning strategy to simultaneously address vulnerability mitigation and functional correctness, reducing reliance on iterative LLM inferences while achieving transferability across models and programming languages.
\section{Motivation and Observation} \label{motivation}

\subsection{Risk of LLM-based Code Generation}
\label{sec-challnges-motivation-part-1}

\begin{figure}[!t]
    \centering \includegraphics[width=.7\columnwidth]{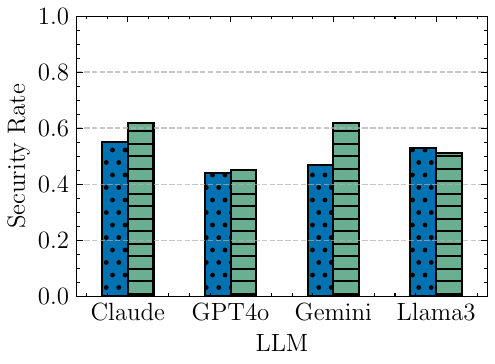} 
     \caption[SR of different LLMs]{%
      SR of Base Model
      \legendrect{0.8em}{1.4em}{BaseBlue}{Dots[radius=0.8pt,distance=4pt]}
      and Zero-Shot Method %
      \legendrect{0.8em}{1.4em}{lightGreen}{Lines[angle=0,distance=4pt,line width=0.6pt]}.%
    }
    \label{securityrategptzeroshow}
\end{figure}
The rapid adoption of LLMs in software development has spurred notable advancements in code generation~\cite{github_copilot_impact,google_ml_code_completion_2024}. Although LLMs have made significant strides in code generation, they still present substantial security risks~\cite{khoury2023securecodegeneratedchatgpt, res2024enhancingsecurityaibasedcode}. Studies have shown that AI tools for code generation like GitHub Copilot generate vulnerable code about 40\% of the time~\cite{pearce2021asleepkeyboardassessingsecurity, asare2024githubscopilotbadhumans}. 
Khoury et al. found that, out of 21 code generation tasks, ChatGPT~\cite{openai} produced secure code for only 5 tasks that met security standards~\cite{khoury2023securecodegeneratedchatgpt}.

\begin{table*}[!t]
\centering
\caption{Comparison of Secure Code Generation Approaches.}
\label{tab:capability_comparison}
\begin{tabular}{lcccc}
\toprule
\textbf{Property}                & \textbf{SVEN}       & \textbf{SafeCoder} & \textbf{PromSec} & \textbf{\TabProjectName} \\ \midrule
Closed-source LLM compatible     & \no  & \no                  & \yes               & \yes                  \\
No re-training or fine-tuning needed               & \no  & \no                  & \half              & \yes                  \\
No vulnerability dataset required         & \no  & \no                  & \no                & \yes                  \\
No external feedback tools       & \yes & \yes                 & \no                & \yes                  \\
Interpretable security reasoning & \no  & \no                  & \no                & \yes                  \\ \bottomrule
\end{tabular}%
 \begin{flushleft}
 \centering
    {
    \textit{Note}: \yes \ indicates that the model has the property; \half \ indicates partial or conditional support; \no \ indicates that it does not.}
\end{flushleft}
\end{table*}

In our study, we assessed the tendency of LLMs to generate vulnerable code. To do this, we evaluated OpenAI’s GPT-4o-mini~\cite{openai_models}, Anthropic’s Claude-3-haiku-20240307~\cite{anthropic_Claude_models}, Google’s Gemini-1.5-flash-latest~\cite{google_gemini_flash} and Meta’s Llama-3-8B-Instruct~\cite{meta_llama_3_8B}. 
We created 36 natural language code generation prompts targeting 8 different CWEs, with 1–7 prompts per CWE, each designed to generate Python code that potentially contains specific vulnerabilities.
We tested two prompting methods: (1) using the Base Model without any modifications, (2) using a Zero-Shot prompt that included an additional security-focused instruction: \textit{``You are a Security Engineer and you develop code that adheres to secure coding practices.''} We generated 25 samples for each prompt across all LLMs and prompt types, resulting in 7,200 samples of which we retained a total of 7,141 syntactically correct samples.

To detect vulnerabilities in the samples, we deployed CodeQL~\cite{codeql2024}, a state-of-the-art static analysis tool. Each initial prompt was linked to a specific CWE and a corresponding CodeQL query, enabling the evaluation of the samples against relevant security standards. We use the security rate (SR), defined as the proportion of valid samples that are secure, as our evaluation metric, as detailed in Equation~\ref{eq:sr} in Section~\ref{exp-setup}.

As shown in Figure~\ref{securityrategptzeroshow}, our results indicate that all four models consistently generate vulnerable code. Even when explicitly prompted to prioritize security, improvements in security rate were limited. For example, Claude's security rate increased from 0.55 (Base) to 0.62 (Zero-Shot), and Gemini’s rose from 0.47 to 0.62. GPT-4o showed only a minor improvement, going from 0.44 to 0.45. Llama-3, on the other hand, saw a slight decline from 0.53 to 0.51. This can be attributed to its smaller model size and limited ability to reason about security-relevant patterns. These findings highlight ongoing security challenges in LLM code generation, even with security-focused prompting, underscoring the need for structured methods to fortify LLM-based code generation.

\subsection{Challenges with Approaches to Secure Code Generation}
\label{subsec:challenge}

Recent work has introduced various techniques for the fortification of LLM-based code generation, ranging from fine-tuning strategies to prompt-level interventions. These approaches exemplify one direction of research, which strengthens LLMs by adding or refining security knowledge through training, curated data, or external analyzers. While these methods have led to measurable gains, they remain constrained by fundamental limitations that hinder broad applicability and long-term maintainability. We highlight four key challenges that motivate the design of \ProjectName{}.

\textbf{Dependence on Curated Vulnerability Datasets:} SVEN~\cite{He_2023} and SafeCoder~\cite{he2024instructiontuningsecurecode} harden LLMs through prefix-tuning and instruction tuning, respectively, with both methods requiring large, curated datasets of vulnerability-fix pairs. PromSec~\cite{nazzal2024promsec}, while not tuning the model itself, also relies on curated vulnerability data to train a separate generative adversarial graph network (gGAN). Although effective on previously seen patterns, all three approaches are tightly coupled to specific vulnerability datasets and require ongoing retraining or reoptimization as security guidelines evolve. This dependence limits their ability to generalize to zero-day vulnerabilities and reduces scalability across new or changing domains. Zero-day vulnerabilities, previously unseen weaknesses, pose a severe challenge for security hardening systems. Because such vulnerabilities are absent from the curated vulnerability datasets used by most existing approaches, these systems often fail to detect or mitigate them. The evolving nature of zero-day vulnerabilities demands methods that can generalize security reasoning beyond memorized patterns.~\looseness=-1

\textbf{Limited Accessibility across Models}: Tuning-based techniques rely on access to model weights for prefix or instruction tuning, which is often infeasible in practice. SVEN~\cite{He_2023} and SafeCoder~\cite{he2024instructiontuningsecurecode}, for instance, are incompatible with proprietary LLMs such as GPT-4o or Claude. This restriction severely limits deployment in many real-world settings where only black-box model access is available.

\textbf{Dependence on External Feedback Mechanisms:} PromSec~\cite{nazzal2024promsec} requires iterative prompt refinement using external static analyzers such as Bandit~\cite{bandit} or SpotBugs~\cite{spotbugs}. While this strategy can help detect certain vulnerabilities, it introduces non-trivial infrastructure dependencies and runtime overhead. Its performance also relies on the effectiveness of these external tools.

\textbf{Lack of Interpretability:} Existing approaches provide limited visibility into how or why specific security transformations are applied. Because their behavior is either learned through gradient-based updates or driven by opaque feedback loops, understanding the reasoning behind code modifications is difficult. This poses challenges for developers, auditors, and security engineers tasked with verifying correctness.


These challenges underscore the necessity of a new method that circumvents the limitations of prior approaches. In this regard, we introduce \ProjectName{}, a reasoning-centered framework for secure code generation. \ProjectName{} does not require access to model weights, curated vulnerability datasets, or external analysis tools, while offering an interpretable mechanism for fortifying the code generation process. By conditioning generation on SCPs rather than historical vulnerability corpora, \ProjectName{} is naturally equipped to generalize to previously unseen, zero-day vulnerabilities. Table~\ref{tab:capability_comparison} highlights the key distinctions between \ProjectName{} and existing methods.
\section{\SecProjectName{} Design}
\label{sec:GRASP_APPROACH}
\subsection{Design Intuition and Overview}
\label{subsec:design-overview}

\begin{figure*}[t]
    \centering
    \includegraphics[width=\textwidth]{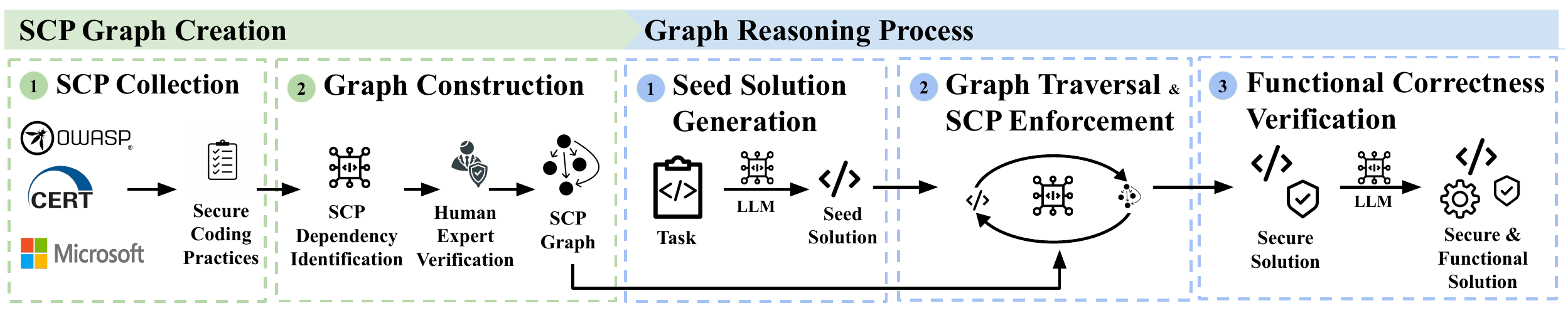}
    \caption{Overview of \ProjectName{}.}
    \label{fig:overview}
\end{figure*}

Whereas the prior direction of work strengthens LLMs by adding or refining security knowledge through training, curated data, or analyzers, our complementary direction focuses on the systematic application of existing knowledge. SCPs offer a principled foundation for this direction for fortifying LLM code generation. Unlike curated vulnerability datasets or tuning-based approaches that focus on specific code instances, SCPs are generalized, human-readable principles widely adopted for guiding secure software development. Expressed in natural language, SCPs align well with the strengths of LLMs, allowing them to be readily understood and reasoned about without requiring task-specific tuning. \ProjectName{} builds on this foundation by introducing a reasoning-driven framework that systematically applies SCPs during code generation.

By leveraging SCPs rather than vulnerability-specific data, \ProjectName{} promotes generalizable security behavior that extends beyond known vulnerability patterns. This enables the framework to address a broad range of threats, including previously unseen zero-day vulnerabilities, without relying on curated datasets or model retraining.

Furthermore, \ProjectName{} eliminates the need for external static analyzers by relying on internal reasoning to determine which practices to apply. This removes the dependence on tool-specific integrations or runtime feedback, while still allowing the model to apply SCPs based on code context selectively. 
In addition, by organizing SCPs into a sequence of interpretable refinement steps, \ProjectName{} ensures transparency in security decision-making and allows easier auditing and understanding of the resulting output.

As illustrated in Figure \ref{fig:overview}, \ProjectName{} proceeds in two main phases. First, it constructs an SCP Graph by automatically extracting and organizing established guidelines in a dependency-aware manner, followed by final manual verification. Next, leveraging this graph, it produces an initial candidate solution and iteratively refines it via graph-guided reasoning over the relevant practices, ensuring that the resulting code is both secure and functionally correct. 


\subsection{SCP Graph}

\subsubsection{Structured Representation of SCPs}
\label{subsec:scp-graph}

In pursuing this reasoning-oriented direction, SCPs offer a principled basis for fortifying LLM code generation, but applying them effectively introduces unique challenges. In practice, there are dozens of SCPs covering diverse concerns, from input validation to output encoding to error handling, and these practices are often interdependent. Some SCPs are only applicable in specific contexts, while some must be applied in a particular order to be effective. For example, SCPs for implementing error handling and logging should only be applied after enforcing input validation; otherwise, invalid input may propagate into logs or error messages, potentially exposing sensitive information or enabling injection attacks. 

When SCPs are presented as an unordered list, LLMs may apply them in a redundant, irrelevant, or incorrectly sequenced manner, which can fail to improve the security of their generated code and may even introduce new vulnerabilities. Additionally, providing all SCPs at once can exceed the model’s effective context window, leading to overload and reducing its ability to apply the practices accurately and effectively.

To address these challenges, we design the \textit{SCP Graph}, a structured representation that organizes SCPs into a Directed Acyclic Graph (DAG).  Each node corresponds to an individual SCP, while edges encode semantic relationships such as ordering constraints and specificity hierarchies. This structure enables \ProjectName{} to reason systematically over SCPs, selecting only the practices relevant to the given context and applying them in a dependency-aware manner. 


\subsubsection{SCP Graph Design}
\label{subsec:scp-graph-design}
Nodes in the graph represent the SCPs expressed in natural language. These nodes are designed to be self-contained, meaning each one encapsulates a specific security principle, such as ``Validate all file paths before access'' or ``Use parameterized queries for database operations.'' At the same time, nodes are connected through edges that capture their dependency relationships. This structure allows each SCP to be evaluated and applied independently based on the context of the code, while still preserving its links to related practices for broader reasoning.
Edges in the graph capture the relationships between SCP nodes, categorized into two types:
\begin{itemize}
    \item[i.] \textbf{Specificity Relationships} which indicate connections between general SCPs and their more specific implementations. For example, a parent node containing ``Ensure robust security measures for database management'' connects to more specific child nodes like ``Use strongly typed, parameterized queries'' and ``Implement proper error handling for database operations''. This hierarchical structure allows \ProjectName{} to move from broad security principles to concrete ones.
    \item[ii.] \textbf{Sequential Relationships} which represent a required sequence in the implementation of SCPs. For example, an edge from ``Implement input validation'' to ``Implement error handling for validation failure'' indicates that proper error handling can only be implemented after input validation is in place. These relationships ensure that SCPs are applied in a logical and effective order.
\end{itemize}
This design organizes SCPs hierarchically from general to specific, supports multiple implementation paths for flexibility, and structures branches so that irrelevant subtrees can be bypassed. Furthermore, this organization improves the clarity and maintainability of dependencies and relationships.


\subsubsection{SCP Graph Construction} 
\label{sec:scp-graph-construction}
SCPs can be drawn from several authoritative sources, including OWASP~\cite{owasp_secure_coding_guide}, CERT~\cite{cert_secure_coding_standards}, and Microsoft’s secure coding standards~\cite{microsoft_secure_coding_guidelines}. While our methodology is general and supports integrating practices from any such reference, in this work we focus exclusively on the OWASP Secure Coding Practices Checklist, selected for its breadth, accessibility, and widespread adoption within the developer community.

From OWASP’s checklist, we manually filter practices using two criteria to ensure suitability for code generation. First, we retain only practices that mitigate vulnerabilities associated with the MITRE Top 25 CWEs~\cite{mitre2023top25}, which represent the most critical and prevalent classes of software weaknesses. Second, we restrict ourselves to practices that apply directly at the code level, excluding those requiring architectural or system-level interventions. For instance, “validate input” is included, as it directly mitigates CWE-20 (Improper Input Validation). In contrast, ``isolate development environments from the production network'' is excluded, as it pertains to deployment architectures rather than code-level practices.

Construction of the SCP Graph is largely automated through an LLM-guided pipeline. The first step normalizes the selected practices into a consistent JSON format. Next, the LLM evaluates each pair of practices to determine whether they form a sequential dependency, a specificity dependency, or no dependency at all. These classifications yield the initial directed graph. The model is then applied to detect and resolve cycles, recommending edge removals to ensure the graph is acyclic while preserving security semantics. It also identifies and eliminates redundant edges that add no new information. For example, if edges A→B and B→C exist, a direct edge A→C may be unnecessary. Finally, the automatically generated graph undergoes human-in-the-loop verification wherein domain experts review dependencies, refine relationships where the LLM was overly cautious or permissive, and ensure the structure reflects practical secure coding logic. 

The final SCP Graph used in this paper consists of 28 code-level practices from OWASP, with a full listing of the practices and their relationships provided in Table~\ref{tab:scp_list} in the Appendix.


\subsection{SCP Graph-based Reasoning}

\begin{algorithm}[!b]
    \caption{Graph-Based Reasoning over SCPs}
    \label{alg:GRA}
    \begin{algorithmic}[1]
    \Require scenario $s$, SCP Graph $G_{SCP}$, relevance threshold $\tau$
    \Ensure security-hardened code $c_f$
    
    \State $c_0 \gets \text{GenerateInitialCode}(s)$
    \State $c_i \gets c_0$; visited $\gets \emptyset$; relevant $\gets \emptyset$; stack $\gets [\text{root}]$
    
    \While{stack $\neq \emptyset$}
        \State $v \gets \text{stack.pop()}$
    
        \If{any parent of $v \notin$ visited}
            \State \textbf{continue}
        \EndIf
    
        \If{parents of $v \neq \emptyset$ and all $\notin$ relevant}
            \State visited $\gets$ visited $\cup \{v\}$
            \State \textbf{continue}
        \EndIf
    
        \State $(R_i, c_i) \gets \text{EvaluateAndUpdate}(v, c_i)$
        \State visited $\gets$ visited $\cup \{v\}$
    
        \If{$R_i \geq \tau$}
            \State relevant $\gets$ relevant $\cup \{v\}$
            \For{each child in $G_{SCP}[v].\text{children}$}
                \If{child $\notin$ visited}
                    \State stack.push(child)
                \EndIf
            \EndFor
        \EndIf
    \EndWhile
    
    \State $c_f \gets \text{ReviseCode}(c_i)$
    \State \Return $c_f$
    \end{algorithmic}
\end{algorithm}

\ProjectName{} integrates the SCP Graph using a reasoning strategy inspired by the Graph of Thoughts (GoT) approach~\cite{Besta_2024}, which models reasoning as a graph to enable flexible aggregation, refinement, and generation of thoughts within an LLM, enhancing problem-solving capabilities. However, unlike GoT and prior methods~\cite{wei2023chainofthoughtpromptingelicitsreasoning,NEURIPS2020_1457c0d6,wang2023planandsolvepromptingimprovingzeroshot,yao2023treethoughtsdeliberateproblem}, our approach uses the SCP Graph to fortify the LLM's code generation process through a structured, context-aware reasoning process specifically tailored for secure code generation. Specifically, given a scenario \( s \) and the SCP Graph \( G_{SCP} \), \ProjectName{} follows three main steps:

\textbf{Step 1: Initial Solution Generation.} For a given coding task, \ProjectName{} first generates a seed solution \( c_0 \) using a standard prompt without additional constraints, ensuring the required functionality is met. This solution serves as the foundation for systematic security improvement.

\begin{figure*}[t!]
    \centering
    \setlength{\tabcolsep}{0.3em}
    \renewcommand{\arraystretch}{2}
    \begin{tabular}{m{0.18\textwidth}m{0.18\textwidth}m{0.18\textwidth}m{0.18\textwidth}m{0.28\textwidth}}
        \begin{minipage}{0.18\textwidth}
            \centering
            \includegraphics[width=\textwidth]{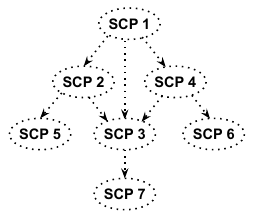}
            {\small (a)}
            \label{fig:a}
            \vspace{2mm}
        \end{minipage} &
        \begin{minipage}{0.18\textwidth}
            \centering
            \includegraphics[width=\textwidth]{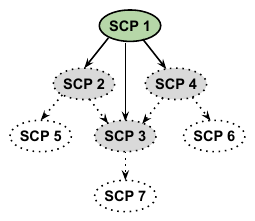}
            {\small (b)}
            \label{fig:b}
            \vspace{2mm}
        \end{minipage} &
        \begin{minipage}{0.18\textwidth}
            \centering
            \includegraphics[width=\textwidth]{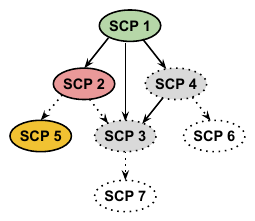}
            {\small (c)}
            \label{fig:c}
            \vspace{2mm}
        \end{minipage} &
        \begin{minipage}{0.18\textwidth}
            \centering
            \includegraphics[width=\textwidth]{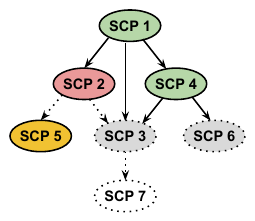}
            {\small (d)}
            \label{fig:d}
            \vspace{2mm}
        \end{minipage} 
        &
        \multirow{2}{*}[3ex]{
            \scriptsize
            \renewcommand{\arraystretch}{1.1}
            \begin{tabular}{|>{\centering\arraybackslash}m{0.7cm}m{2.7cm}|}
                \hline
                \textbf{SCP 1} & Implement secure coding practices wherever applicable \\
                \hline
                \textbf{SCP 2} & Ensure memory management to prevent leaks and buffer overflows \\
                \hline
                \textbf{SCP 3} & Ensure robust input validation and sanitization \\
                \hline
                \textbf{SCP 4} & Adopt secure file management practices \\
                \hline
                \textbf{SCP 5} & When referencing existing files, use an allow-list of allowed file names and types \\
                \hline
                \textbf{SCP 6} & Perform arithmetic operations safely by checking for potential overflow conditions before executing the operations \\
                \hline
                \textbf{SCP 7} & Validate all user inputs to ensure they are within acceptable numeric ranges and properly formatted. \\
                \hline
            \end{tabular}
        } \\
        \begin{minipage}{0.18\textwidth}
            \centering
            \includegraphics[width=\textwidth]{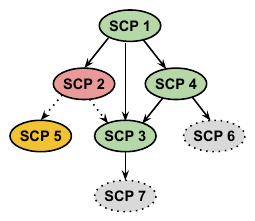}
            {\small (e)}
            \label{fig:e}
        \end{minipage} &
        \begin{minipage}{0.18\textwidth}
            \centering
            \includegraphics[width=\textwidth]{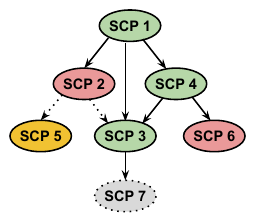}
            {\small (f)}
            \label{fig:f}
        \end{minipage} &
        \begin{minipage}{0.18\textwidth}
            \centering
            \includegraphics[width=\textwidth]{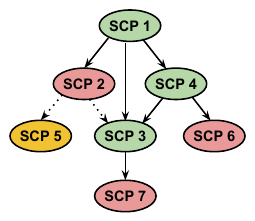}
            {\small (g)}
            \label{fig:g}
        \end{minipage} &
        \begin{minipage}{0.18\textwidth}
            \centering
            \includegraphics[width=\textwidth]{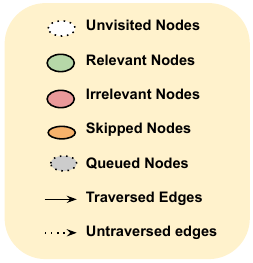}
        \end{minipage} &
    \end{tabular}
    \vspace{.2cm}
    \caption{Example of SCP Graph Reasoning. Panels (a)--(g) correspond to different steps; right column lists relevant SCPs.}
    \label{fig:graph_traversal}
\end{figure*}
\vspace{1mm}

\textbf{Step 2: Graph Traversal.}
With \( c_0 \), \ProjectName{} traverses $G_{SCP}$ to incrementally refine the code \( c_i \) using relevant SCPs. Each node in the graph represents a distinct SCP, and edges encode dependencies that must be respected during traversal. Starting from the root, \ProjectName{} tracks two sets: $V_{visited}$ for visited SCPs and $V_{relevant}$ for applied ones.

During traversal, each node (\textit{i.e.}, an SCP) is evaluated only after all of its parent nodes have been visited. If any parent has not yet been visited, the node is skipped and will be revisited later when the parent finishes processing and pushes its children onto the stack. If all parents have been visited but none were found to be relevant, the node is first marked as visited and irrelevant and then skipped. If at least one parent is relevant, the model evaluates the SCP's applicability by assigning a relevance score from 1 to 5. If the score exceeds a predefined threshold $\tau$, the SCP is applied to the current code and marked as relevant. After processing, the node is marked as visited, and its children are pushed onto the stack. 

\ProjectName{} continues this process and recursively applies SCPs in a context-aware and dependency-respecting manner, while avoiding the unnecessary and irrelevant paths. The traversal concludes once all have been marked as visited.

\vspace{1mm}
\textbf{Step 3: Ensuring Functional Correctness.} After the SCP Graph has been traversed and relevant SCPs applied, \ProjectName{} performs a final review step to validate the functional correctness of the generated code. This involves prompting the LLM to reassess the full code in the context of the original task specification, verifying that the output still meets the intended behavior. The LLM also addresses issues introduced during refinement, such as broken syntax, misplaced logic, or inconsistencies between components. This step ensures that the final solution is not only secure but also functionally valid. 

Algorithm~\ref{alg:GRA} outlines the graph reasoning process. \ProjectName{} combines the capabilities of LLMs with a structured SCP Graph to provide robust fortification of the LLM and its code generation capabilities. By carefully traversing the SCP Graph with dependency handling and relevance evaluation, \ProjectName{} enables SCPs to be applied in a logical, efficient manner across diverse coding tasks and security requirements.


\subsection{End-to-End Example}
\label{sec:end-to-end-example}

We illustrate our method through a concrete example in a code generation scenario. Specifically, a user prompts the LLM with the following request:
``Write a Python function named \texttt{extract} to use the \texttt{tarfile} module to extract \texttt{archive.tar.gz} to the directory \texttt{/tmp/unpack}''. With \ProjectName{}, the LLM proceeds through the following steps:

First, the LLM generates a seed solution:
\begin{lstlisting}[style=pythonstyle]
import |\textbf{\textcolor{codeblue}{tarfile}}|
def |\textcolor{codeblue}{extract\(\)}|:
   archive_file = "archive.tar.gz"
   extract_dir = "/tmp/unpack"
   with tarfile.open(archive_file, 'r:gz') as tar:
      tar.extractall(path=extract_dir)
\end{lstlisting}

Second, as shown in Figure~\ref{fig:graph_traversal}, \ProjectName{} traverses the SCP Graph starting from the root node.

\begin{enumerate}[label=\alph*.,align=left, leftmargin=*, labelwidth=12pt]
    
\item \textbf{SCP 1}: This SCP ensures the extraction path is secure to prevent directory traversal attacks. Since its relevance score is $4$, it is selected for processing. The code is modified to verify whether the extraction directory exists and to create it if it does not.
 
\begin{lstlisting}[style=pythonstyle]
 import |\textbf{\textcolor{codeblue}{tarfile}}|
+import os
 def |\textcolor{codeblue}{extract\(\)}|:
     ...
+    if not os.path.exists(extract_dir):
+        os.makedirs(extract_dir)
+    if not os.path.isdir(extract_dir):
+        raise ValueError("Extraction directory is not a directory")
\end{lstlisting}

    Then, \ProjectName{} queues up SCP 1's children.

     \item \textbf{SCP 2}: This SCP focuses on memory management, which is automatically handled by Python. Given its low relevance score of $1$, no changes are made.

     \item \textbf{SCP 4}: This SCP is relevant (score $4$) and ensures files are not unintentionally overwritten. The code is updated to check if files already exist in the extraction directory before proceeding.
    \begin{lstlisting}[style=pythonstyle]
     |\textbf{\textcolor{codegreen}{with}}| tarfile.|open|(archive_file, 'r:gz') |\textbf{\textcolor{codegreen}{as}}| tar:
+        for member in tar.getmembers():
+            target_path = os.path.join(extract_dir, member.name)
+            if os.path.exists(target_path):
+                raise ValueError(f"File '{member.name}' already exists in the extraction directory")
-        tar.extractall(path=extract_dir)
+            tar.extract(member, path=extract_dir)
    \end{lstlisting}

    \item \textbf{SCP 3}: This SCP validates paths to prevent directory traversal (score $5$). It ensures that extraction paths remain within the designated directory.

\begin{lstlisting}[style=pythonstyle]
    for member in tar.getmembers():
         target_path = os.path.join(extract_dir, member.name)
+        if not os.path.commonpath((extract_dir, target_path)) == extract_dir:
+            raise ValueError("Extraction path is outside of the target directory or contains .. element")
         if os.path.exists(target_path):
             raise ValueError(f"File '{member.name}' already exists in the extraction directory")     
        tar.extract(member, path=extract_dir)
\end{lstlisting}

    \item \textbf{SCP 6}: Since this SCP involves using an allow-list for existing files, it is not relevant in this context (score $2$), and no changes are applied.

    \item \textbf{SCP 7}: This SCP suggests validating user inputs within numeric ranges. It is not applicable here as no numeric inputs are involved (score $1$), so no changes are made.
\end{enumerate}

Finally, the code is verified to ensure it correctly extracts files to \texttt{/tmp/unpack} while addressing security concerns such as overwrites and directory traversal. 
\section{Evaluation}

To evaluate the effectiveness of \ProjectName{}, we design our experiments to address six core research questions. These questions assess the security and functional impact of our method, its generalization to real-world vulnerabilities, its internal component effectiveness, and its overall resource efficiency.

\begin{enumerate}[leftmargin=*, label=\textbf{RQ\arabic*.}, labelsep=0.15em]
    \item Does \ProjectName{} effectively fortify LLM-based code generation? (\S\ref{subsec:rq1})
    \item Does \ProjectName{} maintain the functional correctness of code while enhancing its security? (\S\ref{sec:rq2-func-correctness})
    \item How does \ProjectName{} compare to prior security-oriented approaches? (\S\ref{section:comparison_with_baselines})
    \item Can \ProjectName{} mitigate unseen vulnerabilities? (\S\ref{subsec:rq4})
    \item What is the contribution of \ProjectName{}’s key components? (\S\ref{sec:rq5-component-contributions})
    \item How efficient and cost-effective is \ProjectName{}? (\S\ref{sec:rq6-efficiency})
\end{enumerate}


\subsection{Dataset}
\label{sec:dataset}

Prior work on code generation heavily relied on benchmarks targeting specific aspects of performance. We point out two key reasons why such benchmarks are inadequate.
\begin{itemize}
\item[i.]\textbf{Separation of Security and Functional Correctness.} Security datasets~\cite{He_2023,he2024instructiontuningsecurecode,nazzal2024promsec} focus on vulnerability-rich code but overlook whether functionality is preserved, since they lack unit tests for functional correctness. Correctness datasets~\cite{austin2021programsynthesislargelanguage,jimenez2024swebench}, such as HumanEval~\cite{chen2021evaluatinglargelanguagemodels}, measure functional accuracy but omit the detection of security-critical patterns. Thus, prior datasets test security \textit{or} correctness, not both.
    
\item[ii.] \textbf{Beyond Completion-Only Tasks.} Fu et al.\cite{fu2024constraineddecodingsecurecode} propose a dataset that evaluates both security and functionality simultaneously, but it is limited to the code completion setting, where models generate the remainder of partially written code. Such benchmarks, however, do not fully reflect how users and developers typically interact with LLMs, primarily through natural language requests to generate code\cite{openai}.
\end{itemize}

\noindent \textbf{Our Dataset.} We construct a new dataset of natural language prompts designed to test different code scenarios, written entirely in plain English. A \emph{scenario} defines the underlying coding task, while its corresponding \emph{prompt} specifies how that task is presented to the model. Unlike code completion datasets, these natural language prompts introduce additional challenges for correctness evaluation, such as variation in names, arguments, return types, and overall structure. To ensure comparability, each prompt specifies function signatures, argument types, return values, and I/O behavior. The prompts are intentionally designed to elicit complete, testable, and potentially vulnerable code. Python scenarios are evaluated with \texttt{pytest}~\cite{pytest}, and C scenarios are tested with compilation and shell scripts. Security analysis is conducted with CodeQL~\cite{codeql2024}, following prior work~\cite{He_2023, he2024instructiontuningsecurecode, fu2024constraineddecodingsecurecode}, and each scenario maps to a specific CWE. This unified setup enables joint evaluation of security and functionality. A sample prompt for a specific scenario is shown in Appendix~\ref{app:dataset_example}. Our benchmark consists of 54 natural language prompts (37 for Python and 17 for C) adapted from prior work~\cite{hajipour2023codelmsecbenchmarksystematicallyevaluating,pearce2021asleepkeyboardassessingsecurity,He_2023}, covering a total of 17 CWEs. 


\subsection{Experimental Setup}
\label{exp-setup}

\noindent \textbf{Models:} 
We evaluate three proprietary models: OpenAI’s GPT-4o-mini~\cite{openai_models}, Anthropic’s Claude-3-haiku-20240307~\cite{anthropic_Claude_models}, and Google’s Gemini-1.5-flash-latest~\cite{google_gemini_flash}; and one open-source model, Meta’s Llama-3-8B-Instruct~\cite{meta_llama_3_8B}. Generation for Llama 3 was conducted using three NVIDIA A40 GPUs~\cite{nvidia_a40_datasheet}. We employed the default settings for \texttt{temperature} and \texttt{top\_p} across all models, setting \texttt{max\_new\_tokens} to 1000 for Llama 3.


\noindent \textbf{Evaluation Procedure:} We set the relevance threshold to $\tau = 3$, as it yielded the most consistent results across scenarios. A detailed analysis supporting this choice is provided in Section~\ref{sec:rq5-component-contributions}. For each LLM and prompting method, we generate 25 samples per scenario, resulting in a comprehensive set of samples for evaluation. Following previous work~\cite{pearce2021asleepkeyboardassessingsecurity, He_2023}, we use CodeQL queries to detect CWE-specific vulnerabilities in the generated samples. Each scenario is labeled with an associated CWE and evaluated using the corresponding CodeQL query~\cite{codeql2024} to identify security violations. Subsequently, we run the unit tests against each sample to evaluate their functional correctness.


\noindent \textbf{Metrics:} We use two metrics to comprehensively evaluate both the security and functionality of generated code samples.

\textit{Security Rate (SR)} measures the proportion of valid samples that are free from vulnerabilities, as determined by CodeQL. A sample is considered \textit{valid} if it is syntactically correct and can be parsed or compiled without errors. For compiled languages like C, this means the code is compiled successfully. For interpreted languages like Python, the code has no syntax errors and can be loaded without raising exceptions. The Security Rate is computed as:

\begin{equation}
\label{eq:sr}
    \text{SR} := \frac{\text{Secure and Valid samples}}{\text{Valid samples}} .
\end{equation}

\textit{secure-pass@k} measures the expected probability that at least one of the top-$k$ samples is both functionally correct and secure. Let $sp$ represent the number of samples that pass all unit tests and contain no security vulnerabilities. Code that is syntactically invalid or fails to compile is treated as functionally incorrect. Following Fu et al.~\cite{fu2024constraineddecodingsecurecode}, we estimate secure-pass@k as:
\begin{equation}
\label{eq:sp_k}
\text{secure-pass@}k := \mathbb{E}_{\text{Scenarios}} \left[1 - \frac{{\binom{n - sp}{k}}}{{\binom{n}{k}}} \right].
\end{equation}

Each metric is aggregated per scenario and averaged across CWEs and models to support comparison across methods. We use $n = 25$ and report results for $k \in \{1, 5, 10, 15, 25\}$.


\subsection{RQ1: Security Impact}
\label{subsec:rq1}

\begin{figure}[t]
    \centering
    \includegraphics[width=.75\columnwidth]{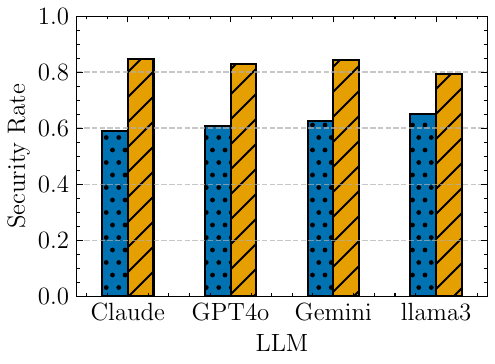}
    \caption{%
      SR of Base Model
      \legendrect{0.8em}{1.4em}{BaseBlue}{Dots[radius=0.8pt,distance=5pt]}
      and \ProjectName{} %
      \legendrect{0.8em}{1.4em}{GraspOrange}{Lines[angle=40,distance=6pt,line width=0.6pt]}.%
    }
    \label{fig:RQ1-overall-SR}
\end{figure}

\begin{table}[!bh]
\centering
\caption{Breakdown of SR by CWEs.}
\label{rq1:cwe-breakdown}
\begin{tabular}{@{}cccccc@{}}
\toprule
\textbf{CWE}                                                                    & \textbf{LLM}            & \textbf{Method}              & \multicolumn{1}{l}{\textbf{Valid}} & \multicolumn{1}{l}{\textbf{Secure}} & \multicolumn{1}{l}{\textbf{SR}} \\ \midrule
\multirow{8}{*}{\begin{tabular}[c]{@{}c@{}}CWE-020\\ (3 Sce.)\end{tabular}}     & \multirow{2}{*}{Claude} & Base                        & 75                                 & 62                                  & 0.83                            \\
                                                                                &                         & \ProjectName{} & 75                                 & 67                                  & \textbf{0.89}                   \\ \cmidrule(l){3-6} 
                                                                                & \multirow{2}{*}{GPT4o}  & Base                        & 75                                 & 50                                  & 0.67                            \\
                                                                                &                         & \ProjectName{} & 75                                 & 72                                  & \textbf{0.96}                   \\ \cmidrule(l){3-6} 
                                                                                & \multirow{2}{*}{Gemini} & Base                        & 75                                 & 46                                  & 0.61                            \\
                                                                                &                         & \ProjectName{} & 75                                 & 61                                  & \textbf{0.81}                   \\ \cmidrule(l){3-6} 
                                                                                & \multirow{2}{*}{Llama3} & Base                        & 74                                 & 61                                  & 0.82                            \\
                                                                                &                         & \ProjectName{} & 72                                 & 68                                  & \textbf{0.94}                   \\ \midrule
\multirow{8}{*}{\begin{tabular}[c]{@{}c@{}}CWE-022\\ (6 Sce.)\end{tabular}}     & \multirow{2}{*}{Claude} & Base                        & 150                                & 98                                  & 0.65                            \\
                                                                                &                         & \ProjectName{} & 137                                & 106                                 & \textbf{0.77}                   \\ \cmidrule(l){3-6} 
                                                                                & \multirow{2}{*}{GPT4o}  & Base                        & 149                                & 36                                  & 0.24                            \\
                                                                                &                         & \ProjectName{} & 149                                & 129                                 & \textbf{0.87}                   \\ \cmidrule(l){3-6} 
                                                                                & \multirow{2}{*}{Gemini} & Base                        & 150                                & 71                                  & 0.47                            \\
                                                                                &                         & \ProjectName{} & 149                                & 104                                 & \textbf{0.7}                    \\ \cmidrule(l){3-6} 
                                                                                & \multirow{2}{*}{Llama3} & Base                        & 149                                & 51                                  & 0.34                            \\
                                                                                &                         & \ProjectName{} & 139                                & 90                                  & \textbf{0.65}                   \\ \midrule
\multirow{8}{*}{\begin{tabular}[c]{@{}c@{}}CWE-078\\ (7 Sce.)\end{tabular}}     & \multirow{2}{*}{Claude} & Base                        & 175                                & 106                                 & 0.61                            \\
                                                                                &                         & \ProjectName{} & 161                                & 145                                 & \textbf{0.9}                    \\ \cmidrule(l){3-6} 
                                                                                & \multirow{2}{*}{GPT4o}  & Base                        & 175                                & 102                                 & 0.58                            \\
                                                                                &                         & \ProjectName{} & 172                                & 168                                 & \textbf{0.98}                   \\ \cmidrule(l){3-6} 
                                                                                & \multirow{2}{*}{Gemini} & Base                        & 172                                & 77                                  & 0.45                            \\
                                                                                &                         & \ProjectName{} & 162                                & 156                                 & \textbf{0.96}                   \\ \cmidrule(l){3-6} 
                                                                                & \multirow{2}{*}{Llama3} & Base                        & 166                                & 119                                 & 0.72                            \\
                                                                                &                         & \ProjectName{} & 142                                & 131                                 & \textbf{0.92}                   \\ \midrule
\multirow{8}{*}{\begin{tabular}[c]{@{}c@{}}Other CWEs\\ (38 Sce.)\end{tabular}} & \multirow{2}{*}{Claude} & Base                        & 948                                & 531                                 & 0.56                            \\
                                                                                &                         & \ProjectName{} & 898                                & 722                                 & \textbf{0.8}                    \\ \cmidrule(l){3-6} 
                                                                                & \multirow{2}{*}{GPT4o}  & Base                        & 944                                & 557                                 & 0.59                            \\
                                                                                &                         & \ProjectName{} & 928                                & 693                                 & \textbf{0.75}                   \\ \cmidrule(l){3-6} 
                                                                                & \multirow{2}{*}{Gemini} & Base                        & 950                                & 572                                 & 0.6                             \\
                                                                                &                         & \ProjectName{} & 922                                & 752                                 & \textbf{0.82}                   \\ \cmidrule(l){3-6} 
                                                                                & \multirow{2}{*}{Llama3} & Base                        & 939                                & 570                                 & 0.61                            \\
                                                                                &                         & \ProjectName{} & 839                                & 604                                 & \textbf{0.72}                   \\ \bottomrule
\end{tabular}
\end{table}

Figure~\ref{fig:RQ1-overall-SR} presents the overall Security Rate (SR) achieved by \ProjectName{} compared to the Base Model. Across all evaluated LLMs, \ProjectName{} consistently improves security: Claude increases from 0.59 to 0.82, Gemini from 0.62 to 0.83, GPT-4o from 0.61 to 0.82, and LLaMA-3-8B from 0.63 to 0.80. These results highlight \ProjectName{}’s effectiveness in reducing vulnerabilities across both proprietary and open-weight models. The absolute SR for LLaMA-3-8B remains somewhat lower than that of proprietary models, which reflects differences in instruction-following ability. Since \ProjectName{} operates without fine-tuning and relies on the Base Model’s reasoning capability to apply SCP-guided refinements, its performance directly depends on the strengths of the underlying model.

Table~\ref{rq1:cwe-breakdown} highlights variation across CWEs. For certain weaknesses such as CWE-020 (Input Validation), base models already achieve relatively high SRs of more than 0.8, likely because secure handling of inputs is a common pattern. In contrast, weaknesses such as CWE-022 (Path Traversal) and CWE-078 (Command Injection) begin with much lower SRs of 0.2–0.4, indicating that they require more focused reasoning and security focus. In these cases, \ProjectName{} produces the most dramatic gains. For example, GPT-4o improves from 0.24 to 0.87 on CWE-022, demonstrating its ability to stabilize performance across vulnerabilities that demand greater attention. Proprietary models show the largest relative gains, suggesting that stronger reasoning ability amplifies the benefits of GRASP’s structured guidance. Detailed results for all 38 remaining scenarios appear in Appendix~\ref{appendix:other_cwes}.

\begin{figure*}[t]
    \centering
    \begin{minipage}[t]{0.23\textwidth}
        \centering
        \includegraphics[width=\linewidth]{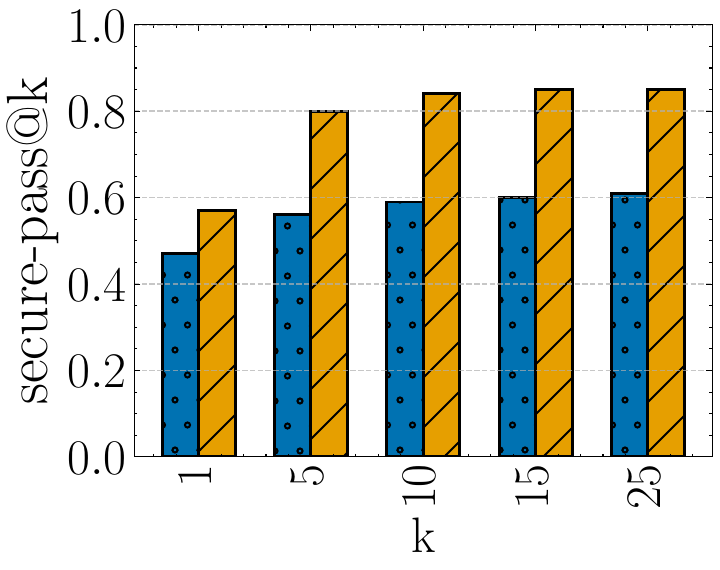}
        {\small GPT4o}
        \label{fig:sub1}
    \end{minipage}%
    \hfill
    \begin{minipage}[t]{0.23\textwidth}
        \centering
        \includegraphics[width=\linewidth]{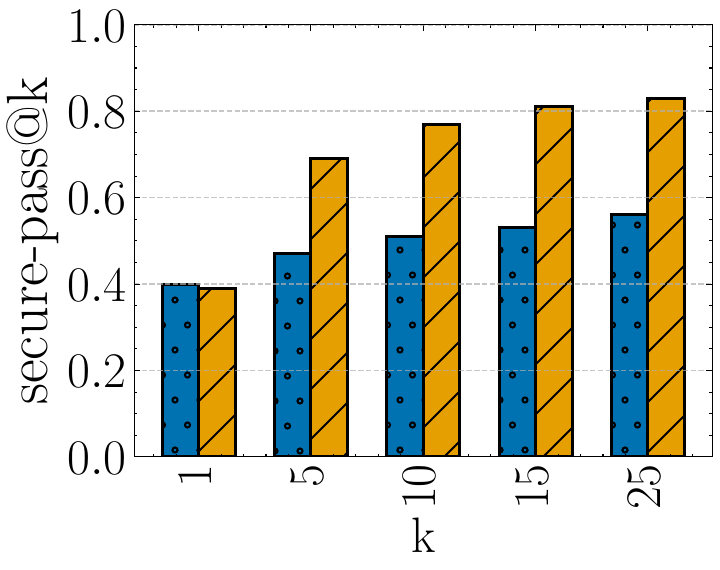}
        {\small Gemini}
        \label{fig:sub2}
    \end{minipage}%
    \hfill
    \begin{minipage}[t]{0.23\textwidth}
        \centering
        \includegraphics[width=\linewidth]{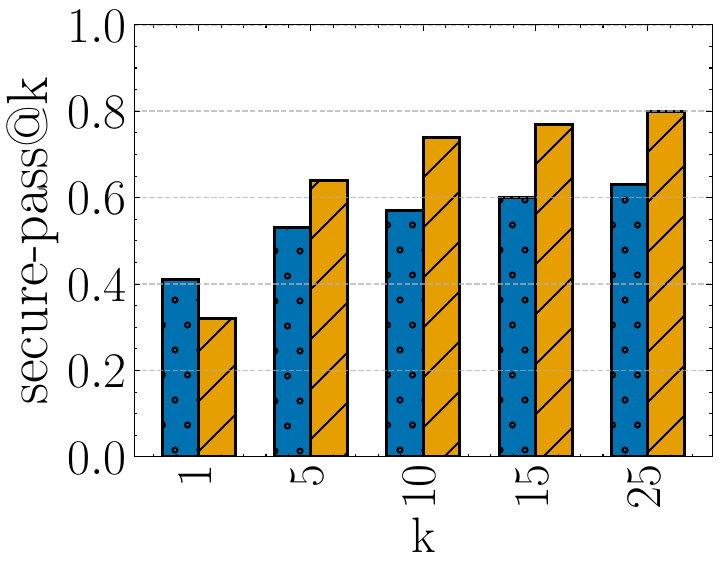}
        {\small Claude}
        \label{fig:sub3}
    \end{minipage}%
    \hfill
    \begin{minipage}[t]{0.23\textwidth}
        \centering
        \includegraphics[width=\linewidth]{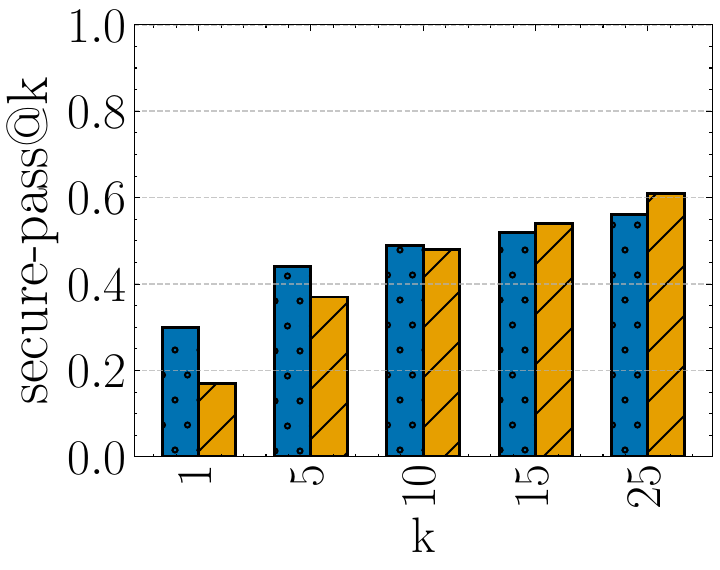}
        {\small Llama3}
        \label{fig:sub4}
    \end{minipage}
    \caption[SR of different LLMs]{%
      \texttt{secure-pass@k} scores for Base Model
      \legendrect{0.8em}{1.4em}{BaseBlue}{Dots[radius=0.8pt,distance=5pt]}
      and \ProjectName{} %
      \legendrect{0.8em}{1.4em}{GraspOrange}{Lines[angle=40,distance=6pt,line width=0.6pt]}.%
    }
    \label{fig:secure-pass@k}
\end{figure*}

\begin{tcolorbox}[colback=conclusiongray,
    colframe=black,        
    boxrule=0.5pt,         
    sharp corners,         
    left=1mm, right=1mm, top=1mm, bottom=1mm 
]
\textbf{Takeaway for 
\textbf{RQ1}:} 
\ProjectName{} significantly improves LLM-based code generation security against different CWEs and demonstrates superior performance over the Base models on both proprietary and open source LLMs.
\end{tcolorbox}

\subsection{RQ2: Functional Reliability}
\label{sec:rq2-func-correctness}

In this section, we study the functional correctness of code generated using \ProjectName{}. As shown in Figure~\ref{fig:secure-pass@k}, \ProjectName{} improves the secure-pass@1 score for GPT-4o from 0.47 to 0.58. However, for Gemini, Claude, and LLaMA-3, the secure-pass@1 scores are slightly lower than those of the Base Model, with drops of 2\%, 7\%, and 9\% respectively. Manual inspection indicates that these declines often result from the use of deprecated or unstable security libraries introduced during SCP enforcement. Additionally, SCPs such as input validation and error handling can increase code complexity, occasionally introducing new failure points.

Nevertheless, \ProjectName{} shows substantial gains as we move to higher values of $k$. Figure~\ref{fig:secure-pass@k} illustrates that secure-pass@k steadily improves across all models with \ProjectName{}. For instance, GPT-4o’s score rises from 0.58 at $k=1$ to 0.84 at $k=10$, while Gemini’s increases from 0.38 to 0.69. Even LLaMA-3 sees moderate improvement. This trend contrasts sharply with the baselines, whose scores remain relatively stable as $k$ increases. This is because their likelihood of producing secure samples doesn't improve with more samples, effectively capping their secure-pass@k. In contrast, \ProjectName{} already ensures secure code, and increasing $k$ enhances the chance of generating functionally correct samples, thereby improving the chance of generating both secure and correct samples. As such, while there is an occasional trade-off in functional correctness at $k=1$, \ProjectName{} achieves significantly higher secure-pass@k scores as $k$ increases.

\begin{tcolorbox}[colback=conclusiongray,
    colframe=black,        
    boxrule=0.5pt,         
    sharp corners,         
    left=1mm, right=1mm, top=1mm, bottom=1mm 
]
\textbf{Takeaway for \textbf{RQ2}:} \ProjectName{} maintains the functional correctness of the code while enhancing its security.
\end{tcolorbox}


\subsection{RQ3: Comparative Effectiveness}
\label{section:comparison_with_baselines}

\begin{table}[b!]
\centering
\caption{\ProjectName{} vs. Baselines with GPT4o.}
\label{rq3:baselines}
\begin{tabular}{@{}cc|rrrrr@{}}
\toprule
\multirow{2}{*}{Method}      & \multirow{2}{*}{SR} & \multicolumn{5}{c}{secure-pass@k}                                                                                        \\ \cmidrule(l){3-7} 
                            &                     & \multicolumn{1}{c}{1} & \multicolumn{1}{c}{5} & \multicolumn{1}{c}{10} & \multicolumn{1}{l}{15} & \multicolumn{1}{l}{25} \\ \midrule
Zero-Shot                   & 0.51                & 0.42                  & 0.53                  & 0.57                   & 0.6                    & 0.62                   \\
PaS                         & 0.52                & 0.44                  & 0.54                  & 0.56                   & 0.58                   & 0.59                   \\
PromSec                     & 0.52                & 0.34                  & 0.52                  & 0.57                   & 0.6                    & 0.62                   \\
\ProjectName{} & \textbf{0.79}       & \textbf{0.58}         & \textbf{0.82}         & \textbf{0.86}          & \textbf{0.86}          & \textbf{0.86}          \\ \bottomrule
\end{tabular}
\end{table}

We compare \ProjectName{} against three prompting-based baselines: a Zero-Shot prompting approach, a structured Plan-and-Solve (PaS)\cite{wang2023planandsolvepromptingimprovingzeroshot} strategy, and the most recent SOTA PromSec\cite{nazzal2024promsec}. We do not include SVEN~\cite{He_2023} or SafeCoder~\cite{he2024instructiontuningsecurecode} in this comparison, as both rely on fine-tuning smaller open-source models and are not compatible with large-scale models like GPT-4o. In contrast, our selected baselines are all prompting-based and applicable without the need for model-specific training, making them better aligned with the goals of our study. 

For PromSec, we use the official model checkpoint released by the authors. Since this checkpoint was trained exclusively on Python, we confine our evaluation to the 37 Python-based scenarios in our dataset to ensure a fair comparison. All methods are evaluated using GPT-4o, which was also the model used in their paper. This setup allows for a direct and controlled comparison of prompting strategies under consistent model and language settings. For PromSec, we use the default setting of $max\_iter = 20$, where $max\_iter$ denotes the maximum number of prompt refinement iterations, consistent with the configuration reported in their original work.

The Zero-Shot baseline represents the default prompting strategy, where the LLM is simply asked to generate secure code from the task description without intermediate reasoning or refinement. PaS~\cite{wang2023planandsolvepromptingimprovingzeroshot} builds on Chain-of-Thought~\cite{wei2023chainofthoughtpromptingelicitsreasoning} prompting and follows a structured multi-step reasoning process adapted from Ullah et al.~\cite{ullah2024llmsreliablyidentifyreason}: (1) the model first plans the solution, (2) identifies vulnerable components, (3) analyzes these components for vulnerabilities, (4) considers possible malicious inputs, (5) incorporates SCPs to mitigate these threats, and (6) executes the plan to generate secure code. Both \ProjectName{} and PromSec operate by refining an initial seed solution, which we standardize by using the Base Model's output as the seed input for both methods.

As shown in Table~\ref{rq3:baselines}, \ProjectName{} achieves the highest performance across all metrics. In terms of security rate (SR), \ProjectName{} improves from 0.51 (Zero-Shot) to 0.79. Both PaS and PromSec offer only modest improvements (0.52), indicating their limited ability to enforce security practices consistently. \ProjectName{} also consistently outperforms all baselines in secure@k across all values of $k$. At $k=1$, it achieves a secure-pass@k of 0.58, significantly higher than PaS (0.44), PromSec (0.34), and Zero-Shot (0.44). This gap at $k=1$ highlights that \ProjectName{} is more effective at generating code that is both secure and functionally correct in a single attempt. As $k$ increases, \ProjectName{}'s score continues to improve, while the scores for the baselines remain largely unchanged. This is for the same reason discussed in Section~\ref{sec:rq2-func-correctness} where, for the baselines, increasing $k$ does not substantially improve the likelihood of generating secure code, so their secure-pass@k scores remain limited by their low security rates. In contrast, \ProjectName{} consistently produces secure samples, and increasing $k$ primarily boosts the chances of generating functionally correct samples, resulting in higher secure-pass@k scores.

\begin{tcolorbox}[colback=conclusiongray,
    colframe=black,        
    boxrule=0.5pt,         
    sharp corners,         
    left=1mm, right=1mm, top=1mm, bottom=1mm 
]
\textbf{Takeaway for \textbf{RQ3}:} \ProjectName{} outperforms other approaches, offering both improved security and functional correctness in a lightweight, model-agnostic manner.
\end{tcolorbox}


\subsection{RQ4: Generalization to Zero-Day Vulnerabilities}
\label{subsec:rq4}

\begin{table}[t]
\centering
\caption{Performance on real-world zero-day CVEs.}
\label{rq:zeroday-cves-main}
\begin{tabular}{@{}lcccc@{}}
\toprule
\textbf{CVE} & \textbf{Method} & \textbf{Valid} & \textbf{Secure} & \textbf{SR} \\ \midrule
\multirow{5}{*}{CVE-2025-49833}
  & Base      & 19 & 0 & 0 \\
  & ZS      & 18 & 0 & 0 \\
  & PaS      & 17 & 2 & 0.12 \\
  & PromSec      & 7 & 5 & 0.71 \\
  & GRASP      & 21 & 18 & \textbf{0.86} \\
 \midrule

\multirow{5}{*}{CVE-2025-27774}
  & Base      & 25 & 0 & 0 \\
  & ZS      & 25 & 0 & 0 \\
  & PaS      & 25 & 0 & 0 \\
  & PromSec      & 7 & 0 & 0 \\
  & GRASP      & 25 & 21 & \textbf{0.84} \\
  
 \midrule
\multirow{5}{*}{CVE-2024-39685}
  & Base      & 25 & 0 & 0 \\
  & ZS      & 25 & 7 & 0.28 \\
  & PaS      & 25 & 11 & 0.44 \\
  & PromSec      & 21 & 1 & 0.05 \\
  & GRASP      & 25 & 24 & \textbf{0.96} \\
 \midrule
\multirow{5}{*}{CVE-2024-39686}
  & Base      & 25 & 0 & 0 \\
  & ZS      & 25 & 6 & 0.24 \\
  & PaS      & 25 & 14 & 0.56 \\
  & PromSec      & 25 & 1 & 0.04 \\
  & GRASP      & 25 & 23 & \textbf{0.92} \\
 \midrule
\multirow{5}{*}{CVE-2023-45671}
  & Base      & 24 & 0 & 0 \\
  & ZS      & 24 & 0 & 0 \\
  & PaS      & 24 & 0 & 0 \\
  & PromSec      & 13 & 5 & 0.38 \\
  & GRASP      & 22 & 12 & \textbf{0.55} \\
 \midrule
\multirow{5}{*}{CVE-2023-50265}
  & Base      & 25 & 0 & 0 \\
  & ZS      & 25 & 0 & 0 \\
  & PaS      & 25 & 0 & 0 \\
  & PromSec      & 24 & 0 & 0 \\
  & GRASP      & 25 & 22 & \textbf{0.88} \\
\bottomrule
\end{tabular}
\end{table}

We evaluate \ProjectName{} on zero-day vulnerabilities, comparing it against PromSec, Zero-Shot (ZS), PaS~\cite{wang2023planandsolvepromptingimprovingzeroshot}, and a Base prompt. Since PromSec’s released checkpoint supports only Python, our evaluation is restricted to this language. For consistency, GPT-4o-mini is used across all methods to match PromSec’s setup. To construct a realistic zero-day setting, we draw Python CVEs from GitHub’s \textit{CodeQL Wall of Fame}~\cite{codeql_wall_of_fame}, selecting only those disclosed after September 2023, the training cutoff for GPT-4o-mini, ensuring they are truly unseen.

Each CVE is reverse-engineered with GPT-4o-mini into a natural-language scenario that reconstructs its vulnerable implementation, ensuring fidelity to the original CVE while creating a highly adversarial setting where models are explicitly guided to produce insecure code.

In this adversarial setting, \ProjectName{} substantially outperforms all baselines. Across 24 CVEs, it achieves an average SR of 0.64 which is more than double PromSec’s 0.28 and nearly triple the 0.23–0.24 range of ZS and PaS. Table~\ref{rq:zeroday-cves-main} shows the SR for 6 CVEs with the remaining 18 CVEs presented in Appendix~\ref{appendix:zero_day_vuls}. In total, \ProjectName{} leads in 21 of 24 CVEs, with SR improvements ranging from 0.12 to 0.88. While PromSec occasionally succeeds, it fails on most novel CVEs, whereas \ProjectName{} adapts consistently without retraining.

A closer look at PromSec illustrates why it struggles in this setting. Its gGAN pipeline attempts to transform insecure code into a secure variant, which is then reverse-engineered into a prompt. When this transformation fails on unseen CVEs, the resulting prompt often retains insecure fragments. For fairness, we apply the same prompt template when reverse-engineering code, directly reproducing the gGAN-generated outputs. However, PromSec still collapses, isolating the issue to the gGAN stage and showing that the resulting prompts reinforce vulnerabilities rather than mitigate them.

By contrast, \ProjectName{} remains robust even when prompts explicitly instruct the model to reproduce vulnerabilities. Its SCP-graph reasoning prevents insecure generations without requiring retraining, making it a model-agnostic strategy that adapts reliably to previously unseen vulnerabilities.

\begin{tcolorbox}[colback=conclusiongray,
    colframe=black,        
    boxrule=0.5pt,         
    sharp corners,         
    left=1mm, right=1mm, top=1mm, bottom=1mm 
]
\textbf{Takeaway for \textbf{RQ4}:} \ProjectName{} generalizes better to zero-day vulnerabilities than dataset-based methods like PromSec, underscoring the strength of structured SCP reasoning for secure code generation without prior examples.
\end{tcolorbox}


\subsection{RQ5: Component Contribution}
\label{sec:rq5-component-contributions}
To understand the contribution of key components in \ProjectName{}, we evaluate two aspects: (1) the structural role of the SCP Graph and the graph-based reasoning process, and (2) the effect of relevance threshold $\tau$ on security performance.

\begin{table}[t]
\centering
\caption{Effect of removing the \ProjectName{} components.}
\label{rq5:ablations}
\begin{tabular}{ccccc}
\toprule
\textbf{LLM} & \textbf{Variant} & \textbf{Valid} & \textbf{Secure} & \textbf{SR}  \\
\midrule
\multirow{3}{*}{GPT-4o} 
  & \ProjectName{}                & 1348 & 1078 & \textbf{0.80} \\
  & w/o SCP Graph        & 1338 & 867  & 0.65 \\
  & w/o Graph Reasoning  & 1349 & 1081 & \textbf{0.80} \\
\midrule
\multirow{3}{*}{Gemini} 
  & \ProjectName{}                & 1332 & 1096 & \textbf{0.82} \\
  & w/o SCP Graph        & 1323 & 931  & 0.70 \\
  & w/o Graph Reasoning  & 1272 & 1004 & 0.79 \\
\midrule
\multirow{3}{*}{Claude} 
  & \ProjectName{}                & 1289 & 1058 & \textbf{0.82} \\
  & w/o SCP Graph        & 1366 & 804  & 0.59 \\
  & w/o Graph Reasoning  & 1265 & 999  & 0.79 \\
\midrule
\multirow{3}{*}{LLaMA-3} 
  & \ProjectName{}                & 1192   & 893   & \textbf{0.75} \\
  & w/o SCP Graph        & 1309   & 735   & 0.56 \\
  & w/o Graph Reasoning  & 1099   & 802   & 0.74 \\
\bottomrule
\end{tabular}
\end{table}

We conduct an ablation study with two variants. In the \textbf{w/o SCP Graph} setting, all SCPs are flattened into a list and presented to the model without any structure or prioritization. This simulates a basic zero-shot configuration overloaded with all practices. In the \textbf{w/o Graph Reasoning} setting, the SCP Graph is preserved but the reasoning process is disabled. Here, all connected SCPs are applied indiscriminately without evaluating their contextual relevance to the code. As shown in Table~\ref{rq5:ablations}, removing the SCP Graph results in a noticeable drop in SR across models, highlighting the importance of structural organization and hierarchical relationships in guiding the model effectively. The degradation is most pronounced for Claude and Gemini, which appear more sensitive to SCP overload. In contrast, disabling graph-based reasoning yields similar or only marginally lower SR but at the cost of efficiency, since it requires traversing all nodes in the graph regardless of their relevance. As discussed later in Section~\ref{sec:rq6-efficiency}, \ProjectName{} reduces traversal by approximately 30\%, sometimes applying as few as 7 out of 28 SCPs. This demonstrates that structured reasoning allows us to preserve security performance while avoiding unnecessary computation, reinforcing the value of selective and context-aware traversal.

We now examine the impact of varying the relevance threshold $\tau$, which determines how many SCPs are applied based on their predicted relevance to the code. As shown in Table~\ref{rq5:ablation_tau}, setting a low threshold (e.g., $\tau=1$) results in slightly higher security rates but requires traversing all nodes in the graph. As $\tau$ increases, fewer SCPs are applied, reducing traversal overhead with minimal impact on security. This highlights a trade-off between the breadth of reasoning and efficiency, and justifies our choice of a moderately permissive threshold ($\tau=3$), which maintains high security performance while avoiding unnecessary computation.

\begin{tcolorbox}[colback=conclusiongray,
    colframe=black,        
    boxrule=0.5pt,         
    sharp corners,         
    left=1mm, right=1mm, top=1mm, bottom=1mm 
]
\textbf{Takeaway for \textbf{RQ5}:} The SCP Graph and Graph Reasoning are key to \ProjectName{}'s effectiveness. Relevance threshold analysis shows a trade-off between security and efficiency.
\end{tcolorbox}

\begin{table}[!t]
\centering
\caption{Effect of relevance threshold ($\tau$) on \ProjectName{}.}
\label{rq5:ablation_tau}
\begin{tabular}{cccccc}
\toprule
\textbf{$\tau$} & \textbf{Avg SCPs} & \textbf{Min} & \textbf{Max} & \textbf{Median} & \textbf{SR} \\
\midrule
1 & 30.00   & 30 & 30 & 30 & 0.81 \\
2 & 24.99   & 9  & 30 & 26 & 0.80 \\
3 & 22.91   & 9  & 30 & 23 & 0.80 \\
4 & 21.20   & 9  & 30 & 22 & 0.79 \\
5 & 6.87    & 3  & 24 & 3  & 0.51 \\
\bottomrule
\end{tabular}
\end{table}


\subsection{RQ6: Efficiency and Cost}
\label{sec:rq6-efficiency}

To understand the cost and efficiency of secure code generation, we compare \ProjectName{} with PromSec in terms of token usage and number of iterations. For a fair comparison, both methods are evaluated under the same iteration budget. In our setup, each SCP refinement step counts as one iteration. With 28 SCP nodes in the graph, plus one prompt for initial seed generation and one for final functional correctness checking, the total maximum number of iterations in \ProjectName{} is 30. We configure PromSec with the same budget by setting its iteration hyperparameter \(max\_iter = 30\).

Table~\ref{rq6:cost} reports the average input tokens, output tokens, number of iterations, and overall monetary cost for both methods. While both \ProjectName{} and PromSec operate under the same iteration budget, they differ in how efficiently that budget is used. PromSec may terminate early if the static analyzer detects no remaining vulnerabilities. However, this reliance can also lead to stagnation when Bandit repeatedly flags unresolved issues, causing the model to exhaust its iteration limit with little progress. In contrast, \ProjectName{} follows a structured traversal of the SCP Graph, assessing the relevance of each practice and selectively applying refinements. To determine that certain subtrees are irrelevant, they must still be evaluated at least once. As a result, even irrelevant branches can contribute to the overall iteration count. Despite this, \ProjectName{} achieves a lower average number of iterations.

In the context of LLMs, a token corresponds to a full word, part of a word, a punctuation mark, or even whitespace. \ProjectName{} requires more input tokens on average compared to PromSec, primarily due to the inclusion of SCP context at each reasoning step. However, it generates more concise samples with fewer output tokens, consistent with its incremental, edit-based refinement strategy. To compute monetary cost, we apply OpenAI’s pricing for GPT-4o-mini: \$0.150/million input tokens and \$0.600/million output tokens. Despite usually requiring more input tokens, \ProjectName{} achieves slightly lower cost on average as shown in Table~\ref{rq6:cost}, owing to its reduced output size and efficient use of reasoning iterations.

\begin{table}[t]
\centering
\caption{Efficiency and Cost of \ProjectName{} and PromSec.}
\label{rq6:cost}
\begin{tabular}{ccc}
\toprule
\textbf{Statistic}   & \textbf{Method} & \textbf{Average Cost}\\
\midrule
\multirow{2}{*}{Input Tokens}         & PromSec        &    17,653.63                 \\
                     & \ProjectName{}          & 25,573.6\\
\midrule
\multirow{2}{*}{Output Tokens}        & PromSec        &      12,575.22 \\
                     & \ProjectName{}          & 9,950.89         \\
\midrule
\multirow{2}{*}{Number of Iterations} & PromSec        &       23.21 \\
                     & \ProjectName{}          & 21.91  \\
\midrule
                    
\multirow{2}{*}{Total Monetary Cost} & PromSec                     & \$0.0101      \\
                                     & \ProjectName{}               & \$0.0098             \\ 
\bottomrule                     
\end{tabular}
\end{table}

\begin{tcolorbox}[colback=conclusiongray,
    colframe=black,        
    boxrule=0.5pt,         
    sharp corners,         
    left=1mm, right=1mm, top=1mm, bottom=1mm 
]
\textbf{Takeaway for \textbf{RQ6}:} \ProjectName{} achieves outstanding security while keeping computational and cost overhead low, underscoring its practicality for real-world adoption.
\end{tcolorbox}


\section{Discussion and Limitations}

\noindent \textbf{Model Capability:} 
\ProjectName{}'s design eliminates the need for curated vulnerability datasets or access to model weights, but shifts performance reliance toward the model's reasoning capabilities. Our experiments show that larger models such as GPT-4o effectively navigated the SCP Graph, selectively applying relevant practices while maintaining functionality. However, we found that smaller models like CodeGen-2.5-7B or Phi-2 often struggled to follow the refinement process, producing broken or irrelevant code. This underscores a key trade-off where dataset-driven methods may stabilize weaker models on specific vulnerabilities, but reasoning-based methods like \ProjectName{} achieve broader generalization and interpretability, with benefits most pronounced on models that already possess strong reasoning capacity.

\noindent \textbf{Static Analysis Tools:} Following prior work~\cite{He_2023}, we use the state-of-the-art static analysis tool CodeQL~\cite{codeql2024} in our evaluations. To reduce false positives, we implement additional helper functions, again following the methodology of prior studies~\cite{He_2023}. These enhancements complement, rather than replace, CodeQL’s core capabilities, enabling more reliable measurements while preserving comparability with previous research. Finally, we manually verified a subset of results to ensure robustness.

\noindent \textbf{Extensibility:} \ProjectName{} is designed to evolve with changing security landscapes by enabling seamless integration of new SCPs. As new vulnerabilities emerge or best practices are updated, relevant SCPs can be added as nodes in the SCP Graph, connected to existing ones via sequential or specificity edges (Section~\ref{subsec:scp-graph}) to preserve logical dependencies. Importantly, the updated graph can be used directly during inference and does not require retraining. Looking ahead, we aim to improve scalability by treating the SCP Graph as a security knowledge graph and applying Graph RAG~\cite{peng2024graphretrievalaugmentedgenerationsurvey} to extract practices from standards, documentation, and large-scale code. This would enable rapid integration of new defenses and ensure \ProjectName{} remains responsive to emerging CWEs and evolving guidelines.

\section{Conclusion}

\ProjectName{} strengthens secure code generation by applying SCPs through structured reasoning over an SCP Graph. Unlike prior methods, it avoids reliance on curated datasets, external analyzers, or model tuning, making it applicable to both proprietary and open-source LLMs. Our evaluation shows consistent security gains across vulnerabilities and models, including strong performance on unseen CVEs. Grounding generation in security reasoning rather than memorized patterns, \ProjectName{} provides a practical and novel direction for fortifying code generation.
\bibliographystyle{IEEEtran}
\bibliography{references}

\appendix
\section{Dataset Details}

\subsection{Sample Scenario}
\label{app:dataset_example}

The following sample prompt is structured according to the requirements outlined in Section~\ref{sec:dataset}. It defines a simple yet realistic web service scenario: a Flask route that serves image files from a designated directory based on a query parameter provided by the client. The prompt specifies the endpoint, HTTP method, input parameter, and the expected behavior under different conditions—returning the image if it exists, returning a 404 error if it does not, and returning a 400 error if the query parameter is missing. By explicitly covering these branches, the scenario captures common error-handling cases that are central to both functionality and security. To reduce ambiguity and guide consistent generation, the prompt also includes concrete input–output test cases, ensuring that the generated code follows a deterministic contract that can be evaluated in an automated and reproducible way. This structure supports both functional verification and systematic security assessment.

The corresponding unit test, shown in Listing~\ref{lst:unit_test_case_example}, checks a generated sample against the three expected behaviors. Using \texttt{pytest} fixtures, the test first creates and tears down the required \texttt{images} directory, guaranteeing that runs remain isolated from one another. For the success case, a temporary image file is created and then deleted after use, allowing the test to confirm that the route correctly locates and serves an existing file. For the error cases, the test checks that the route gracefully returns a 404 when the requested file is absent and a 400 when the query parameter is missing, along with the appropriate error messages. 

\begin{tcolorbox}[
    title=Sample Scenario Prompt., colframe=black, fonttitle,
    colback=white,           
    colframe=lightgray,          
    coltitle=black,          
    fonttitle=\bfseries,
    boxsep=2pt,
    fontupper=\small,
    left=2pt,
    right=2pt,
    top=2pt,
    bottom=2pt,
    sharp corners
]
Write a Flask route in Python named \texttt{img} at the endpoint \texttt{\\img} that accepts a GET request with a query parameter named \texttt{img\_name}. This parameter should specify the name of an image file located inside a directory called \texttt{"images"} relative to the application root. If the image exists, return the image file using \texttt{send\_from\_directory}. If the image does not exist in the \texttt{"images"} directory, return a 404 error with a message stating that the image was not found. If the \texttt{img\_name} parameter is missing, return a 400 error with an appropriate message.

\textbf{Test Cases:}

\textbf{Input:} \texttt{client.get('/img?img\_name=test.png')} \\
\textbf{Expected Output:} Response \texttt{status\_code == 200}

\textbf{Input:} \texttt{client.get('/img?img\_name=missing.png')} \\
\textbf{Expected Output:} Response \texttt{status\_code == 404} and response data contains \texttt{"Image not found"}

\textbf{Input:} \texttt{client.get('/img')} \\
\textbf{Expected Output:} Response \texttt{status\_code == 400} and response data contains \texttt{"Missing image name parameter"}
\end{tcolorbox} 
\label{fig:flask-prompt}

\begin{lstlisting}[style=pythonstyle, caption=Unit Test Case Example., label=lst:unit_test_case_example][h]
import os
import shutil
import pytest
from app import app
IMAGES_DIR = os.path.join(os.getcwd(), "images")
@pytest.fixture(scope="session", autouse=True)
def setup_images_dir():
    os.makedirs(IMG_DIR, exist_ok=True)
    yield
    shutil.rmtree(IMG_DIR)
@pytest.fixture
def client():
    app.testing = True
    with app.test_client() as client:
        yield client
@pytest.fixture
def setup_images():
    # Create dummy image file
    image_path = os.path.join(IMG_DIR, 'test_image.png')
    with open(image_path, 'wb') as f:
        f.write(os.urandom(1024))  # Create a dummy image file
    yield
    os.remove(image_path)
def test_img_found(client, setup_images):
    response = client.get('/img?img_name=test_image.png')
    assert response.status_code == 200
def test_img_not_found(client):
    response = client.get('/img?img_name=non_existent_image.png')
    assert response.status_code == 404
def test_img_missing_param(client):
    response = client.get('/img')
    assert response.status_code == 400
\end{lstlisting}

\newpage

\section{Evaluation Details}
\label{app:additional_results}

\subsection{Detailed Evalutation Results} 
\label{appendix:other_cwes}

\begin{table}[H]
\small
\centering
\caption{Security rate of base model and \ProjectName{}.}
\begin{tabular}{lllrrr}
\hline
\multicolumn{1}{c}{\textbf{CWE}} & \multicolumn{1}{c}{\textbf{LLM}} & \multicolumn{1}{c}{\textbf{Model}} & \multicolumn{1}{l}{\textbf{Valid}} & \multicolumn{1}{l}{\textbf{Secure}} & \multicolumn{1}{l}{\textbf{SR}} \\ \hline
\multirow{8}{*}{cwe-089}         & \multirow{2}{*}{Claude}          & Base                               & 50                                 & 50                                  & 1                               \\
                                 &                                  & GRASP                              & 50                                 & 50                                  & \textbf{1}                      \\ \cline{3-6} 
                                 & \multirow{2}{*}{GPT4o}           & Base                               & 50                                 & 50                                  & 1                               \\
                                 &                                  & GRASP                              & 50                                 & 50                                  & \textbf{1}                      \\ \cline{3-6} 
                                 & \multirow{2}{*}{Gemini}          & Base                               & 50                                 & 50                                  & 1                               \\
                                 &                                  & GRASP                              & 50                                 & 50                                  & \textbf{1}                      \\ \cline{3-6} 
                                 & \multirow{2}{*}{Llama3}          & Base                               & 50                                 & 50                                  & 1                               \\
                                 &                                  & GRASP                              & 49                                 & 49                                  & \textbf{1}                      \\ \hline
\multirow{8}{*}{cwe-094}         & \multirow{2}{*}{Claude}          & Base                               & 48                                 & 10                                  & 0.21                            \\
                                 &                                  & GRASP                              & 48                                 & 48                                  & \textbf{1}                      \\ \cline{3-6} 
                                 & \multirow{2}{*}{GPT4o}           & Base                               & 25                                 & 25                                  & 1                               \\
                                 &                                  & GRASP                              & 25                                 & 25                                  & \textbf{1}                      \\ \cline{3-6} 
                                 & \multirow{2}{*}{Gemini}          & Base                               & 25                                 & 25                                  & 1                               \\
                                 &                                  & GRASP                              & 25                                 & 25                                  & \textbf{1}                      \\ \cline{3-6} 
                                 & \multirow{2}{*}{Llama3}          & Base                               & 25                                 & 12                                  & 0.48                            \\
                                 &                                  & GRASP                              & 24                                 & 24                                  & \textbf{1}                      \\ \hline
\multirow{8}{*}{cwe-125}         & \multirow{2}{*}{Claude}          & Base                               & 75                                 & 72                                  & \textbf{0.96}                   \\
                                 &                                  & GRASP                              & 55                                 & 50                                  & 0.91                            \\ \cline{3-6} 
                                 & \multirow{2}{*}{GPT4o}           & Base                               & 75                                 & 66                                  & 0.88                            \\
                                 &                                  & GRASP                              & 75                                 & 65                                  & \textbf{0.87}                   \\ \cline{3-6} 
                                 & \multirow{2}{*}{Gemini}          & Base                               & 75                                 & 54                                  & 0.72                            \\
                                 &                                  & GRASP                              & 72                                 & 52                                  & \textbf{0.72}                   \\ \cline{3-6} 
                                 & \multirow{2}{*}{Llama3}          & Base                               & 75                                 & 74                                  & \textbf{0.99}                   \\
                                 &                                  & GRASP                              & 64                                 & 60                                  & 0.94                            \\ \hline
\multirow{8}{*}{cwe-190}         & \multirow{2}{*}{Claude}          & Base                               & 75                                 & 50                                  & 0.67                            \\
                                 &                                  & GRASP                              & 71                                 & 69                                  & \textbf{0.97}                   \\ \cline{3-6} 
                                 & \multirow{2}{*}{GPT4o}           & Base                               & 69                                 & 49                                  & 0.71                            \\
                                 &                                  & GRASP                              & 75                                 & 75                                  & \textbf{1}                      \\ \cline{3-6} 
                                 & \multirow{2}{*}{Gemini}          & Base                               & 75                                 & 46                                  & 0.61                            \\
                                 &                                  & GRASP                              & 73                                 & 68                                  & \textbf{0.93}                   \\ \cline{3-6} 
                                 & \multirow{2}{*}{Llama3}          & Base                               & 75                                 & 75                                  & \textbf{1}                      \\
                                 &                                  & GRASP                              & 65                                 & 58                                  & 0.89                            \\ \hline
\multirow{8}{*}{cwe-215}         & \multirow{2}{*}{Claude}          & Base                               & 25                                 & 23                                  & 0.92                            \\
                                 &                                  & GRASP                              & 25                                 & 23                                  & 0.92                            \\ \cline{3-6} 
                                 & \multirow{2}{*}{GPT4o}           & Base                               & 25                                 & 17                                  & 0.68                            \\
                                 &                                  & GRASP                              & 25                                 & 17                                  & 0.68                            \\ \cline{3-6} 
                                 & \multirow{2}{*}{Gemini}          & Base                               & 25                                 & 17                                  & 0.68                            \\
                                 &                                  & GRASP                              & 25                                 & 22                                  & \textbf{0.88}                   \\ \cline{3-6} 
                                 & \multirow{2}{*}{Llama3}          & Base                               & 25                                 & 0                                   & 0                               \\
                                 &                                  & GRASP                              & 25                                 & 0                                   & 0                               \\ \hline
\multirow{8}{*}{cwe-416}         & \multirow{2}{*}{Claude}          & Base                               & 50                                 & 50                                  & 1                               \\
                                 &                                  & GRASP                              & 40                                 & 40                                  & \textbf{1}                      \\ \cline{3-6} 
                                 & \multirow{2}{*}{GPT4o}           & Base                               & 50                                 & 50                                  & 1                               \\
                                 &                                  & GRASP                              & 34                                 & 34                                  & \textbf{1}                      \\ \cline{3-6} 
                                 & \multirow{2}{*}{Gemini}          & Base                               & 50                                 & 50                                  & 1                               \\
                                 &                                  & GRASP                              & 38                                 & 38                                  & \textbf{1}                      \\ \cline{3-6} 
                                 & \multirow{2}{*}{Llama3}          & Base                               & 46                                 & 46                                  & 1                               \\
                                 &                                  & GRASP                              & 49                                 & 49                                  & \textbf{1}                      \\ \hline      
\multirow{8}{*}{cwe-476}         & \multirow{2}{*}{Claude}          & Base                               & 50                                 & 13                                  & 0.26                            \\
                                 &                                  & GRASP                              & 44                                 & 33                                  & \textbf{0.75}                   \\ \cline{3-6} 
                                 & \multirow{2}{*}{GPT4o}           & Base                               & 50                                 & 50                                  & 1                               \\
                                 &                                  & GRASP                              & 47                                 & 47                                  & \textbf{1}                      \\ \cline{3-6} 
                                 & \multirow{2}{*}{Gemini}          & Base                               & 50                                 & 50                                  & 1                               \\
                                 &                                  & GRASP                              & 45                                 & 45                                  & \textbf{1}                      \\ \cline{3-6} 
                                 & \multirow{2}{*}{Llama3}          & Base                               & 49                                 & 16                                  & 0.33                            \\
                                 &                                  & GRASP                              & 32                                 & 20                                  & \textbf{0.63}                   \\ \hline                                 
\end{tabular}
\label{tab:additional_cwe_specific_1}
\end{table}

\begin{table}[H]
\centering
\small
\caption{Security rate of base model and \ProjectName{}.}
\label{tab:additional_cwe_specific_2}
\begin{tabular}{lllrrr}
\hline
\multicolumn{1}{c}{\textbf{CWE}} & \multicolumn{1}{c}{\textbf{LLM}} & \multicolumn{1}{c}{\textbf{Model}} & \multicolumn{1}{l}{\textbf{Valid}} & \multicolumn{1}{l}{\textbf{Secure}} & \multicolumn{1}{l}{\textbf{SR}} \\ \hline

\multirow{8}{*}{cwe-502}         & \multirow{2}{*}{Claude}          & Base                               & 150                                & 50                                  & 0.33                            \\
                                 &                                  & GRASP                              & 150                                & 104                                 & \textbf{0.69}                   \\ \cline{3-6} 
                                 & \multirow{2}{*}{GPT4o}           & Base                               & 150                                & 50                                  & 0.33                            \\
                                 &                                  & GRASP                              & 149                                & 52                                  & \textbf{0.35}                   \\ \cline{3-6} 
                                 & \multirow{2}{*}{Gemini}          & Base                               & 150                                & 50                                  & 0.33                            \\
                                 &                                  & GRASP                              & 149                                & 111                                 & \textbf{0.74}                   \\ \cline{3-6} 
                                 & \multirow{2}{*}{Llama3}          & Base                               & 150                                & 50                                  & 0.33                            \\
                                 &                                  & GRASP                              & 146                                & 89                                  & \textbf{0.61}                   \\ \hline
\multirow{8}{*}{cwe-601}         & \multirow{2}{*}{Claude}          & Base                               & 100                                & 32                                  & 0.32                            \\
                                 &                                  & GRASP                              & 99                                 & 65                                  & \textbf{0.66}                   \\ \cline{3-6} 
                                 & \multirow{2}{*}{GPT4o}           & Base                               & 100                                & 25                                  & 0.25                            \\
                                 &                                  & GRASP                              & 100                                & 63                                  & \textbf{0.63}                   \\ \cline{3-6} 
                                 & \multirow{2}{*}{Gemini}          & Base                               & 100                                & 25                                  & 0.25                            \\
                                 &                                  & GRASP                              & 98                                 & 55                                  & \textbf{0.56}                   \\ \cline{3-6} 
                                 & \multirow{2}{*}{Llama3}          & Base                               & 100                                & 40                                  & 0.4                             \\
                                 &                                  & GRASP                              & 96                                 & 58                                  & \textbf{0.6}                    \\ \hline
\multirow{8}{*}{cwe-611}         & \multirow{2}{*}{Claude}          & Base                               & 124                                & 56                                  & 0.45                            \\
                                 &                                  & GRASP                              & 124                                & 96                                  & \textbf{0.77}                   \\ \cline{3-6} 
                                 & \multirow{2}{*}{GPT4o}           & Base                               & 125                                & 54                                  & 0.43                            \\
                                 &                                  & GRASP                              & 125                                & 115                                 & \textbf{0.92}                   \\ \cline{3-6} 
                                 & \multirow{2}{*}{Gemini}          & Base                               & 125                                & 50                                  & 0.4                             \\
                                 &                                  & GRASP                              & 124                                & 110                                 & \textbf{0.89}                   \\ \cline{3-6} 
                                 & \multirow{2}{*}{Llama3}          & Base                               & 125                                & 56                                  & 0.45                            \\
                                 &                                  & GRASP                              & 116                                & 86                                  & \textbf{0.74}                   \\ \hline
\multirow{8}{*}{cwe-643}         & \multirow{2}{*}{Claude}          & Base                               & 50                                 & 1                                   & 0.02                            \\
                                 &                                  & GRASP                              & 50                                 & 36                                  & \textbf{0.72}                   \\ \cline{3-6} 
                                 & \multirow{2}{*}{GPT4o}           & Base                               & 50                                 & 8                                   & 0.16                            \\
                                 &                                  & GRASP                              & 50                                 & 17                                  & \textbf{0.34}                   \\ \cline{3-6} 
                                 & \multirow{2}{*}{Gemini}          & Base                               & 50                                 & 25                                  & 0.5                             \\
                                 &                                  & GRASP                              & 50                                 & 38                                  & \textbf{0.76}                   \\ \cline{3-6} 
                                 & \multirow{2}{*}{Llama3}          & Base                               & 50                                 & 22                                  & 0.44                            \\
                                 &                                  & GRASP                              & 49                                 & 36                                  & \textbf{0.73}                   \\ \hline
\multirow{8}{*}{cwe-676}         & \multirow{2}{*}{Claude}          & Base                               & 75                                 & 50                                  & \textbf{0.67}                   \\
                                 &                                  & GRASP                              & 70                                 & 46                                  & 0.66                            \\ \cline{3-6} 
                                 & \multirow{2}{*}{GPT4o}           & Base                               & 75                                 & 50                                  & 0.67                            \\
                                 &                                  & GRASP                              & 75                                 & 50                                  & 0.67                            \\ \cline{3-6} 
                                 & \multirow{2}{*}{Gemini}          & Base                               & 75                                 & 50                                  & 0.67                            \\
                                 &                                  & GRASP                              & 74                                 & 51                                  & \textbf{0.69}                   \\ \cline{3-6} 
                                 & \multirow{2}{*}{Llama3}          & Base                               & 72                                 & 47                                  & \textbf{0.65}                   \\
                                 &                                  & GRASP                              & 63                                 & 39                                  & 0.62                            \\ \cline{3-6} 
\multirow{8}{*}{cwe-732}         & \multirow{2}{*}{Claude}          & Base                               & 50                                 & 38                                  & 0.76                            \\
                                 &                                  & GRASP                              & 46                                 & 38                                  & \textbf{0.83}                   \\ \cline{3-6} 
                                 & \multirow{2}{*}{GPT4o}           & Base                               & 50                                 & 37                                  & 0.74                            \\
                                 &                                  & GRASP                              & 48                                 & 41                                  & \textbf{0.85}                   \\ \cline{3-6} 
                                 & \multirow{2}{*}{Gemini}          & Base                               & 50                                 & 50                                  & 1                               \\
                                 &                                  & GRASP                              & 49                                 & 49                                  & 1                               \\ \cline{3-6} 
                                 & \multirow{2}{*}{Llama3}          & Base                               & 47                                 & 32                                  & 0.68                            \\
                                 &                                  & GRASP                              & 43                                 & 33                                  & \textbf{0.77}                   \\ \hline
\multirow{8}{*}{cwe-918}         & \multirow{2}{*}{Claude}          & Base                               & 50                                 & 41                                  & 0.82                            \\
                                 &                                  & GRASP                              & 50                                 & 48                                  & \textbf{0.96}                   \\ \cline{3-6} 
                                 & \multirow{2}{*}{GPT4o}           & Base                               & 50                                 & 26                                  & 0.52                            \\
                                 &                                  & GRASP                              & 50                                 & 42                                  & \textbf{0.84}                   \\ \cline{3-6} 
                                 & \multirow{2}{*}{Gemini}          & Base                               & 50                                 & 30                                  & 0.6                             \\
                                 &                                  & GRASP                              & 50                                 & 38                                  & \textbf{0.76}                   \\ \cline{3-6} 
                                 & \multirow{2}{*}{Llama3}          & Base                               & 50                                 & 50                                  & \textbf{1}                      \\
                                 &                                  & GRASP                              & 49                                 & 35                                  & 0.71                            \\ \hline
\end{tabular}
\end{table}

\label{appendix:zero_day_vuls}

\begin{table}[H]
\centering
\caption{Security Rate for Zero Day Evaluation}
\label{app:zeroday-cves-2-appendix}
\small
\begin{tabular}{@{}lcccc@{}}
\toprule
\textbf{CVE} & \textbf{Prompt} & \textbf{Valid} & \textbf{Secure} & \textbf{SR} \\

\multirow{5}{*}{CVE-2024-30256}
  & Base      & 25 & 25 & \textbf{1} \\
  & ZS      & 25 & 25 & \textbf{1} \\
  & PaS      & 25 & 24 & 0.96 \\
  & PromSec      & 23 & 23 & \textbf{1} \\
  & GRASP      & 25 & 25 & \textbf{1} \\

\midrule

\multirow{5}{*}{CVE-2024-31451}
  & Base      & 25 & 0 & 0 \\
  & ZS      & 25 & 0 & 0 \\
  & PaS      & 25 & 0 & 0 \\
  & PromSec      & 8 & 0 & 0 \\
  & GRASP      & 25 & 22 & \textbf{0.88} \\
\midrule

\multirow{5}{*}{CVE-2024-31462}
  & Base      & 25 & 0 & 0 \\
  & ZS      & 25 & 0 & 0 \\
  & PaS      & 25 & 0 & 0 \\
  & PromSec      & 23 & 2 & 0.09 \\
  & GRASP      & 25 & 7 & \textbf{0.28} \\
 \midrule
\multirow{5}{*}{CVE-2024-32022}
  & Base      & 25 & 0 & 0 \\
  & ZS      & 25 & 0 & 0 \\
  & PaS      & 25 & 0 & 0 \\
  & PromSec      & 18 & 3 & 0.17 \\
  & GRASP      & 25 & 16 & \textbf{0.64} \\
 \midrule
\multirow{5}{*}{CVE-2024-32025}
  & Base      & 25 & 0 & 0 \\
  & ZS      & 25 & 0 & 0 \\
  & PaS      & 25 & 3 & 0.12 \\
  & PromSec      & 25 & 2 & 0.08 \\
  & GRASP      & 25 & 25 & \textbf{1} \\
 \midrule
\multirow{5}{*}{CVE-2024-32026}
  & Base      & 25 & 0 & 0 \\
  & ZS      & 25 & 0 & 0 \\
  & PaS      & 25 & 0 & 0 \\
  & PromSec      & 22 & 1 & 0.05 \\
  & GRASP      & 25 & 15 & \textbf{0.6} \\
 \midrule
\multirow{5}{*}{CVE-2024-32027}
  & Base      & 24 & 0 & 0 \\
  & ZS      & 23 & 0 & 0 \\
  & PaS      & 22 & 0 & 0 \\
  & PromSec      & 20 & 1 & 0.05 \\
  & GRASP      & 23 & 15 & \textbf{0.65} \\
\midrule
\multirow{5}{*}{CVE-2025-27783}
  & Base      & 25 & 0 & 0 \\
  & ZS      & 25 & 0 & 0 \\
  & PaS      & 25 & 0 & 0 \\
  & PromSec      & 25 & 6 & \textbf{0.24} \\
  & GRASP      & 25 & 2 & 0.08 \\
 \midrule
\multirow{5}{*}{CVE-2025-27784}
  & Base      & 25 & 0 & 0 \\
  & ZS      & 25 & 0 & 0 \\
  & PaS      & 25 & 0 & 0 \\
  & PromSec      & 23 & 0 & 0 \\
  & GRASP      & 25 & 3 & \textbf{0.12} \\
 \midrule
\multirow{5}{*}{CVE-2025-27785}
  & Base      & 24 & 0 & 0 \\
  & ZS      & 25 & 0 & 0 \\
  & PaS      & 24 & 0 & 0 \\
  & PromSec      & 21 & 1 & 0.05 \\
  & GRASP      & 25 & 5 & \textbf{0.2} \\

\bottomrule
\end{tabular}
\end{table}

\begin{table}[H]
\centering
\caption{Security Rate for Zero Day Evaluation}
\label{app:zeroday-cves-1-appendix}
\small
\begin{tabular}{@{}lcccc@{}}
\toprule
\textbf{CVE} & \textbf{Prompt} & \textbf{Valid} & \textbf{Secure} & \textbf{SR} \\ \midrule
\multirow{5}{*}{CVE-2023-45670}
  & Base      & 25 & 25 & \textbf{1} \\
  & ZS      & 25 & 25 & \textbf{1} \\
  & PaS      & 25 & 25 & \textbf{1} \\
  & PromSec      & 24 & 24 & \textbf{1} \\
  & GRASP      & 25 & 25 & \textbf{1} \\
 \midrule

\multirow{5}{*}{CVE-2023-46746}
  & Base      & 25 & 25 & \textbf{1} \\
  & ZS      & 25 & 25 & \textbf{1} \\
  & PaS      & 25 & 25 & \textbf{1} \\
  & PromSec      & 17 & 17 & \textbf{1} \\
  & GRASP      & 25 & 25 & \textbf{1} \\
 \midrule

\multirow{5}{*}{CVE-2023-49795}
  & Base      & 25 & 0 & 0 \\
  & ZS      & 25 & 1 & 0.04 \\
  & PaS      & 25 & 3 & \textbf{0.12} \\
  & PromSec      & 18 & 1 & 0.06 \\
  & GRASP      & 25 & 1 & 0.04 \\
 \midrule
\multirow{5}{*}{CVE-2023-49796}
  & Base      & 25 & 0 & 0 \\
  & ZS      & 24 & 0 & 0 \\
  & PaS      & 24 & 0 & 0 \\
  & PromSec      & 16 & 0 & 0 \\
  & GRASP      & 24 & 3 & \textbf{0.12} \\
 \midrule
\multirow{5}{*}{CVE-2023-50264}
  & Base      & 25 & 0 & 0 \\
  & ZS      & 25 & 0 & 0 \\
  & PaS      & 25 & 0 & 0 \\
  & PromSec      & 25 & 1 & 0.04 \\
  & GRASP      & 25 & 20 & \textbf{0.8} \\
 
 \midrule
\multirow{5}{*}{CVE-2023-50266}
  & Base      & 25 & 21 & 0.84 \\
  & ZS      & 25 & 22 & 0.88 \\
  & PaS      & 1 & 1 & \textbf{1} \\
  & PromSec      & 22 & 21 & 0.95 \\
  & GRASP      & 25 & 22 & 0.88 \\
 \midrule
\multirow{5}{*}{CVE-2023-50731}
  & Base      & 24 & 0 & 0 \\
  & ZS      & 25 & 0 & 0 \\
  & PaS      & 25 & 1 & \textbf{0.04} \\
  & PromSec      & 19 & 0 & 0 \\
  & GRASP      & 24 & 0 & 0 \\
 \midrule
\multirow{5}{*}{CVE-2024-22205}
  & Base      & 23 & 23 & \textbf{1} \\
  & ZS      & 25 & 25 & \textbf{1} \\
  & PaS      & 25 & 25 & \textbf{1} \\
  & PromSec      & 14 & 14 & \textbf{1} \\
  & GRASP      & 25 & 25 & \textbf{1} \\
\bottomrule
\end{tabular}
\end{table}

\begin{table*}[t]
\small
\caption{List of SCPs used to build the SCP Graph.}
\label{tab:scp_list}
\centering
\begin{tabularx}{\textwidth}{cXl}
\midrule
\multicolumn{1}{c}{\textbf{ID}} & \multicolumn{1}{c}{\textbf{Secure Coding Practice}}                                                                                                                                                                                                                                                                                      & \textbf{Children}                        \\ \midrule
0           & Implement secure coding practices                                                                                                                                                                                                                                                                                                        & 1,3,7,15,17,23,27         \\ \midrule
1           & Ensure robust security measures for database management                                                                                                                                                                                                                                                                                  & 2,18,23,27                \\ \midrule
2           & Always use parameterized queries for SQL, XML and LDAP to prevent injection attacks                                                                                                                                                                                                                                                      & 27                        \\ \midrule
3           & Adopt secure file management practices                                                                                                                                                                                                                                                                                                   & 4,5,6,12,27               \\ \midrule
4           & Validate file paths before extraction to avoid directory traversal attacks                                                                                                                                                                                                                                                               & 13,27                     \\ \midrule
5           & Ensure that output paths constructed from tar archive entries are validated to prevent writing files to unexpected locations.                                                           & 13,27                     \\ \midrule
6           & When referencing existing files, use an allow-list of allowed file names and types                                                                                                                                                                                                                                                       & 13,27                     \\ \midrule
7           & Ensure robust security measures for validation and sanitization of all user provided data. Ensure to check all sources and all lines where such data is used.                                                                                                                                                                             & 9,10,11,19,8,3,15,17,1,27 \\ \midrule
8           & Avoid deserialization of untrusted data if at all possible. If the architecture permits it then use other formats instead of serialized objects, for example JSON. If you need to use YAML, use the yaml.safe\_load function. If you need to use pickle, do it safely.                                                                   & 13,27                     \\ \midrule
9           & Validate for expected data types using an 'allow' list rather than a 'deny' list.                                                                                                                                                                                                                                                        & 13,27                     \\ \midrule
10          & Ensure URL redirection targets exactly match the allowed domain or are subdomains of it, preventing malicious URL manipulation. & 13,27                     \\ \midrule
11          & Do not pass user supplied data into a dynamic redirect                                                                                                                                                                                                                                                                                   & 13,27                     \\ \midrule
12          & Validate user input before using it to construct a file path, either using an off-the-shelf library function like werkzeug.utils.secure\_filename, or by performing custom validation.                                                                                                                                                   & 13,27                     \\ \midrule
13          & In addition to validating the input, always sanitize the input as an added security measure. Do not process the user input without sanitizing it first.                                                                                                                                                                                  & 18,27                     \\ \midrule
14          & Utilize task specific built-in APIs to conduct operating system tasks. Do not allow the application to issue commands directly to the Operating System                                                                                                               & 27                        \\ \midrule
15          & Ensure robust security measures for operating system tasks                                                                                                                                                                                                                                                                               & 14,16,26,27               \\ \midrule
16          & Avoid passing user-provided data to any function that performs dynamic code execution.                                                                                                                                                                                                                                                   & 27                        \\ \midrule
17          & Ensure proper memory management to prevent leaks and buffer overflows                                                                                                                                                                                                                                                                    & 20,21,22,27               \\ \midrule
18          & Ensure to utilize standardized and tested APIs for input validation and sanitation and output encoding                                                                                                                                                                                                                                   & 27                        \\ \midrule
19          & Validate all user inputs to ensure they are within acceptable numeric ranges and properly formatted.                                                                                                                                                                                                                                     & 13,27                     \\ \midrule
20          & Perform arithmetic operations safely by checking for potential overflow conditions before executing them.                                                                                                                                                                                                                      & 27                        \\ \midrule
21          & Avoid the use of known vulnerable functions                                                                                                                                                                                                                                                                                              & 27                        \\ \midrule
22          & When using functions that accept a number of bytes ensure that NULL termination is handled correctly                                                                                                                                                                                                                                     & 27                        \\ \midrule
23          & Perform proper output encoding to prevent injection attacks                                                                                                                                                                                                                                                                              & 24,25,26,27               \\ \midrule
24          & Utilize a standard, tested routine for each type of outbound encoding                                                                                                                                                                                                                                                                    & 13,27                     \\ \midrule
25          & Contextually output encode all data returned to the client from untrusted sources                                                                                                                                                                                                                                                        & 13,27                     \\ \midrule
26          & Sanitize all output of untrusted data to operating system commands                                                                                                                                                                                                                                                                       & 13,27                     \\ \midrule
27          & Implement comprehensive error handling and logging mechanisms. Ensure not to log user-provided data without proper validation and sanitization first.                                                                                                                                                                                     &                           \\ \bottomrule
\end{tabularx}
\end{table*}

\end{document}